\documentstyle[prd,aps,preprint,amssymb,12pt,epsfig,fleqn]{revtex}

\begin{document}
\draft
\title{
\begin{flushright} \small \rm
  hep-ph/yymmxxx \\ 
  UWThPh-1998-60 \\
  HEPHY-PUB 702/98 \\
  FTUV/98-96 \\ 
  IFIC/98-98 \\ \today
\end{flushright}
More on higher order decays of the lighter top squark}
\author{W.~Porod\thanks{Electronic address: porod@ap.univie.ac.at} \\[0.3cm]
       }
\address{Institut f\"ur Theoretische Physik, Universit\"at Wien,
         A-1090 Vienna, Austria \\[0.3cm] and \\[0.3cm]
        Departamento de F\'\i sica Te\'orica, IFIC-CSIC, 
         Universidad de Valencia, \\  
         Burjassot, Valencia 46100, Spain} 
\maketitle

\begin{abstract}
We discuss the three-body decays 
${\tilde t}_1 \to W^+ \, b \, {\tilde \chi}^0_1$,
${\tilde t}_1 \to H^+ \, b \, {\tilde \chi}^0_1$,
${\tilde t}_1 \to b \, {\tilde l}^+_i \, \nu_l$, and
${\tilde t}_1 \to b \, {\tilde \nu}_l \, l^+$ ($l =e,\mu,\tau$)
of the lighter top squark ${\tilde t}_1$ within the minimal
supersymmetric standard model. We give the complete analytical formulas for
the decay widths and present a numerical study in view of an
upgraded Tevatron, the CERN LHC, and a future lepton collider demonstrating the
importance of these decay modes. 
\end{abstract}
\pacs{14.80.Ly, 11.30.Pb, 12.60.Jv}

\tightenlines

\section{Introduction}

The search for supersymmetry (SUSY) \cite{susy,Haber84} plays an important 
r\^ole
in the experimental program at the colliders LEP2 and Tevatron. It will
be even more important at future 
colliders, e.g. an upgraded Tevatron, LHC, an $e^+ e^-$ linear collider or a
$\mu^+ \mu^-$ collider. Therefore many phenomenological studies have been
carried out in recent years (see e.g. 
\cite{phen1,phen2,muon,epem} and references therein).

Within the supersymmetric extensions of the standard model (SM) the minimal 
supersymmetric standard model (MSSM) \cite{Haber84,Tata95} is the most 
investigated one. The MSSM implies that
every SM fermion has two spin 0 partners called sfermions ${\tilde f}_L$
and ${\tilde f}_R$. In general sfermions decay according to 
${\tilde f}_k \to f \, {\tilde \chi}^0_i, f' \, {\tilde \chi}^\pm_j$ where
${\tilde \chi}^0_i$ and ${\tilde \chi}^\pm_j$ denote neutralinos and charginos,
respectively. 
Here we assume that the lightest neutralino is the lightest supersymmetric
particle (LSP).

Owing to large Yukawa couplings the sfermions of the third generation 
have a quite different phenomenology compared to those of the first two 
generations (see e.g. \cite{Bartl97a} and references therein). The large
Yukawa couplings imply a large mixing between ${\tilde f}_L$ and ${\tilde f}_R$
and large couplings to the higgsino components of neutralinos and charginos.
This is in particular the case for the lighter top squark ${\tilde t}_1$ 
because of the large top quark mass \cite{Ellis83}. The large top quark
mass also implies the existence of
scenarios where all two-body decay modes of ${\tilde t}_1$
(e.g. ${\tilde t}_1 \to t \, {\tilde \chi}^0_i, b \, {\tilde \chi}^+_j,
t \, \tilde g$)  are kinematically forbidden at tree-level. In these
scenarios higher order decays of ${\tilde t}_1$ become relevant:
\cite{Hikasa87,Porod97}:
\begin{eqnarray}
{\tilde t}_1 &\to& c \, {\tilde \chi}^0_{1,2} \\
{\tilde t}_1 &\to& W^+ \, b \, {\tilde \chi}^0_1 \\
{\tilde t}_1 &\to& H^+ \, b \, {\tilde \chi}^0_1 \\
{\tilde t}_1 &\to& b \, {\tilde l}^+_i \, \nu_l \\
{\tilde t}_1 &\to& b \, {\tilde \nu}_l \, l^+ \,, 
\end{eqnarray}
where $l$ denotes $e,\mu,\tau$.
 
In \cite{Hikasa87} it has been shown that decays into sleptons are dominating
over the decays into $ c \, {\tilde \chi}^0_{1,2}$ if they are
 kinematically allowed. However, they have used the approximation: $m_b = 0$, 
$h_b = h_l = 0$ ($l=e,\mu,\tau$),
$m_{ {\tilde t}_1} \ll m_{{\tilde \chi}^+_1} \ll m_{{\tilde \chi}^+_2}$.
In \cite{Porod97} it has been shown that for small $\tan \beta$ the
decay ${\tilde t}_1 \to W^+ \, b \, {\tilde \chi}^0_1$
in general dominates over ${\tilde t}_1 \to c \, {\tilde \chi}^0_{1,2}$ 
whereas for large $\tan \beta$ their branching ratios can be of comparable 
size. 
In this paper we present the complete formulas for the three-body
decays which are so far missing in the literature.
We also perform a numerical analysis for the
mass range of an upgraded Tevatron, the LHC, and a future lepton collider
including the possibility that all of the above decay channels are 
simultaneously open.
In particular it turns out that
the inclusion of the bottom and tau Yukawa couplings 
$h_b$ and $h_\tau$ is important.

This paper is organized as follows:
In Sect.~II we fix our notation and give the analytical expressions for the
decay amplitudes  together with the relevant parts of the MSSM 
Lagrangian. In Sect.~III we present our numerical results for the
branching ratios of the three-body decays in scenarios accessible either at
the Tevatron run II, LHC, or a future lepton collider. Our conclusions
are drawn in Sect.~IV. The analytical formulas for the squared amplitudes
are listed in Appendix \ref{appA}, and Appendix \ref{appB} gives
the various couplings. 

\section{Formulas for the Decay Widths}
\label{sec:threebody}

In this section we fix our notation and we give the Lagrangian relevant 
for the calculation of the decay widths. Moreover, we present the analytical 
formulas  for the matrix elements and generic formulas for the decay widths. 
The complete formulas for the latter are rather lengthy and are listed in 
Appendix \ref{appA}.

The parameters relevant for the following discussion are $M '$, $M$, $m_{A^0}$,
$\mu$, $\tan \beta$, 
$M_{{\tilde D}_i}$, $M_{{\tilde Q}_i}$, $M_{{\tilde U}_i}$, 
$A_{d_i}$, and $A_{u_i}$.
$M'$ and $M$ are the $U(1)$ and $SU(2)$ gaugino masses, for which we
assume the GUT relation $M' = 5/3 \tan^2 \theta_W M$.
$\mu$ is the parameter of the Higgs superpotential, $m_{A^0}$ the mass 
of the pseudoscalar Higgs boson, and $\tan \beta = v_2 / v_1$
where $v_i$ denotes the vacuum expectation value of the Higgs doublet $H_i$.
$M_{{\tilde D}_i}$, $M_{{\tilde Q}_i}$ and $M_{{\tilde U}_i}$
are soft SUSY breaking masses
for the squarks, $A_{d_i}$ and $A_{u_i}$ are trilinear Higgs--squark couplings,
and $i=1,2,3$ is the generation index.

The mass matrix for sfermions in the $(\tilde{f}_L,\:\tilde{f}_R)$ basis
has the following form \cite{Bartl97a}:
\begin{eqnarray} \label{sqm}
  {\cal M}^2_{\tilde{f}_i} = \left(\begin{array}{ll}
                       m_{\tilde{f}_{Li}}^2 & a_{f_i} m_{f_i} \\
                       a_{f_i} m_{f_i}  & m_{\tilde{f}_{Ri}}^2
                   \end{array} \right)
\end{eqnarray}
with
\begin{eqnarray}
m_{\tilde{u}_{Li}}^2&=& M_{{\tilde Q}_i}^2 + m_{u_i}^2 + m_Z^2\cos 2\beta\,
  (\textstyle \frac{1}{2} - \frac2{3} \sin^2\theta_W ) , \nonumber \\
m_{\tilde{u}_{Ri}}^2&=&M_{{\tilde U}_i}^2 + m_{u_i}^2 + \textstyle \frac2{3}
                        m_Z^2\cos 2\beta\,\sin^2\theta_W , \nonumber \\
m_{\tilde{d}_{Li}}^2&=&M_{{\tilde Q}_i}^2 + m_{d_i}^2 - m_Z^2\cos 2\beta\,
    (\textstyle \frac{1}{2} - \frac{1}{3} \sin^2\theta_W ) , \nonumber \\
m_{\tilde{d}_{Ri}}^2&=&M_{{\tilde D}_i}^2 + m_{d_i}^2 -
      \textstyle \frac{1}{3} m_Z^2\cos 2\beta\,\sin^2\theta_W , \nonumber \\
m_{\tilde{l}_{Li}}^2&=&M_{{\tilde L}_i}^2 + m_{l_i}^2 - m_Z^2\cos 2\beta\,
    (\textstyle \frac{1}{2} - \sin^2\theta_W ) , \nonumber \\
m_{\tilde{l}_{Ri}}^2&=&M_{{\tilde E}_i}^2 + m_{l_i}^2 -
      \textstyle m_Z^2\cos 2\beta\,\sin^2\theta_W ,
\label{smatrix}
\end{eqnarray}
and
\begin{eqnarray}
  a_{u_i}m_{u_i} &=& m_{u_i}(A_{u_i} - \mu\cot\beta ), \nonumber \\
  a_{d_i}m_{d_i} &=& m_{d_i}(A_{d_i} - \mu\,\tan\beta ), \nonumber \\
  a_{l_i}m_{l_i} &=& m_{l_i}(A_{l_i} - \mu\,\tan\beta ),
  \label{offdiag}
\end{eqnarray}
where $i$ is a generation index 
($ u_i= u,c,t; \, d_i= d,s,b; \, l_i= e,\nu,\tau$)
which will be suppressed in the following.
The mass eigenstates $\tilde{f}_1$ and $\tilde{f}_2$ are related to
$\tilde{f}_L$ and $\tilde{f}_R$ by:
\begin{eqnarray}
  {\tilde{f}_1 \choose \tilde{f}_2} =
    \left(\begin{array}{ll}
        \cos \theta_{\tilde f} & \sin \theta_{\tilde f} \\
       -\sin \theta_{\tilde f} & \cos \theta_{\tilde f} \end{array} \right)\:
    {\tilde{f}_L \choose \tilde{f}_R} =
    {\cal{R}}^{\tilde f} {\tilde{f}_L \choose \tilde{f}_R} 
  \label{mixing}
\end{eqnarray}
with the eigenvalues
\begin{eqnarray}
  m_{\tilde{f}_{1,2}}^2 = \textstyle \frac{1}{2}\,
       (m_{\tilde{f}_L}^2 + m_{\tilde{f}_R}^2) \mp
         \frac{1}{2} \sqrt{(m_{\tilde{f}_L}^2 - m_{\tilde{f}_R}^2)^2
                           + 4\,a_f^2 m_f^2} \;.
  \label{eq:masses}
\end{eqnarray}
The mixing angle $\theta_{f}$ is given by
\begin{eqnarray}
  \cos \theta_{\tilde f} = \frac{- a_f m_f}
           { \sqrt{(m_{\tilde{f}_L}^2-m_{\tilde{f}_1}^2)^2 + a_f^2 m_f^2}},
  \hspace{5mm}
  \sin \theta_{\tilde f} =
       \sqrt{ \frac{(m_{\tilde{f}_L}^2-m_{\tilde{f}_1}^2)^2}
                   {(m_{\tilde{f}_L}^2-m_{\tilde{f}_1}^2)^2 + a_f^2 m_f^2}} \;.
 \label{eq:mixangl}
\end{eqnarray}
The sneutrino mass is given by:
\begin{equation}
m_{\tilde{\nu}_i}^2 = M_{{\tilde L}_i}^2 + \textstyle \frac{1}{2}
    m_Z^2\cos 2\beta
\end{equation}

The part of the Lagrangian, which is needed for the calculation
of the three-body  decay widths, is given by:
\begin{eqnarray}
{\cal L}_I & = &
 g \sum_{j=1,2}
   \bar{b} \left( k^{\tilde t}_{1j} P_L + l^{\tilde t}_{1j} P_R \right)
         {\tilde \chi}^+_j {\tilde t}_1 +
 g \sum_{ \stackrel{j,n=1,2}{f=e,\mu, \tau} }
    \overline{\nu_f} \left( k^{\tilde f}_{nj} P_L +
         l^{\tilde f}_{nj} P_R \right) {\tilde \chi}^-_j {\tilde f}_n
\nonumber \\ & &
 + g \sum_{ \stackrel{j=1,2}{f=e,\mu, \tau} }
   \overline{{\tilde \chi}^+_j}
      \left( l^{{\tilde \nu}_f}_{j} P_L \right) f \,
          {\tilde \nu}^\dagger_f   
  + g \sum_{\stackrel{k=1,2}{f=b,t}}  \bar f
       \left( b^{f}_{k1} P_L + a^{f}_{k1} P_R \right)
      {\tilde \chi_1^{0}} {\tilde f}_k \nonumber \\
 && - g W_{\mu }^{-} \sum^2_{j=1,2}
\overline{{\tilde \chi}_1^0} \left( O^L_{j1} P_L +
   O^R_{j1} P_R \right) \gamma^{\mu } \tilde \chi_j^+
  -ig \, W^-_\mu
      \sum^2_{j=1,2} A^W_{{\tilde t}_1 {\tilde b}_j}
    \tilde b_j^\dagger
\stackrel{\leftrightarrow }{ \partial_{\mu }} \tilde t_1 \nonumber \\
  && - \frac{g}{\sqrt 2}
    W_{\mu }^- \bar{b} \gamma^{\mu } P_L t 
  - \, g \, H^-  \sum^2_{j=1,2} \overline{{\tilde \chi}^0_1}
               \left( Q^R_{1j}{}' P_L + Q^L_{1i}{}' P_R \right)
               {\tilde \chi}^+_j  
 - \, g  \, H^-  \sum^2_{j=1,2} C^H_{{\tilde t}_1 {\tilde b}_j}
      {\tilde t}_1 \, {\tilde b}^\dagger_j \nonumber \\
 & & + \frac{g}{\sqrt2 \, m_W}
    H^- \bar{b} \left( m_b \tan \beta P_L + m_t \cot \beta P_R \right) t
\label{eq:interaction}
\end{eqnarray}
where $P_{R,L}= ( 1 \pm \gamma_5 ) / 2 $. The various couplings are given
in Appendix~\ref{appB}.

The formula for the decay width
$\Gamma(\tilde t_1 \to W^+ \, b \, \tilde \chi^0_1)$
has already been given in \cite{Porod97}. Therefore, we give only the 
corresponding matrix element
$M_{\tilde t_1 \to W^+ \, b \, \tilde \chi^0_1}$:
\begin{eqnarray}
\hspace*{-7mm}
M_{\tilde t_1 \to W^+ \, b \, \tilde \chi^0_1} & = &  
  - \frac{g^2}{\sqrt2} \sum^2_{j=1} A^W_{{\tilde t}_1 {\tilde b}_j}
     \frac{(p_{\tilde t} + p_{{\tilde b}_j} )^{\mu }}
    {p_{{\tilde b}_j}^2 - m^2_{{\tilde b}_j}
        - i m_{{\tilde b}_j} \Gamma_{\tilde{b}_j} } \bar u(p_b)
     \left[ b^{\tilde{b}}_{j1} P_L + a^{\tilde{b}}_{j1} P_R \right]
                v (p_{{\tilde \chi}^0_1} )
     \epsilon_{\mu} (p_W ) \nonumber \\
 & & + \, g^2  \sum^2_{j=1} \bar u(p_b)
    \left[ l^{\tilde t}_{1j} P_R + k^{\tilde t}_{1j} P_L \right]
   \frac{ \not \! p_{{\tilde \chi}^+_j} - m_{\tilde{\chi}^+_j} }
  { p_{{\tilde \chi}^+_j}^2 - m^2_{\tilde{\chi}^+_j}
  - i m_{\tilde{\chi}^+_j} \Gamma_{\tilde{\chi}^+_j}} \nonumber \\
 & & \hspace{20mm} *  \left[ O^L_{1j}{}' P_L + O^R_{1j}{}' P_R  \right]
    \gamma^{\mu} v(p_{{\tilde \chi}^0_1} ) \epsilon_{\mu} (p_W) \nonumber \\
 & &  - \frac{g^2}{\sqrt2} \bar u (p_b ) \gamma^{\mu } P_L
   \frac{ \not \! p_t +m_t }{ p_{t}^2 -m_{t}^2  - i m_t \Gamma_{t}}
   \left[ b^{\tilde{t}}_{11} P_L + a^{\tilde{t}}_{11} P_R \right]
      v (p_{{\tilde \chi}^0_1} ) \epsilon_{\mu } (p_W )
\label{eq:TfistbWchi}
\end{eqnarray}
In \cite{Porod97} also the formula for 
$\Gamma(\tilde t_1 \to c \, \tilde \chi^0_1)$
\cite{Hikasa87} has been rewritten in the notation used here. 

\begin{figure}
 \unitlength 1mm
 \begin{picture}(135,175)
   \put(43,153){\large $\tilde{t}_1$}
   \put(106,144){\large $\tilde{\chi}^0_1$}
   \put(65,161){\large $H^+$}
   \put(76,153){\large $\tilde{b}_{1,2}$}
   \put(106,155){\large $b$}
   \put(43,93){\large $\tilde{t}_1$}
   \put(108,98){\large $H^+$}
   \put(107,83){\large $\tilde{\chi}^0_1$}
   \put(75,94){\large $\tilde{\chi}^-_{1,2}$}
   \put(68,101){\large $b$}
   \put(43,33){\large $\tilde{t}_1$}
   \put(65,42){\large $\tilde{\chi}^0_1$}
   \put(108,37){\large $H^+$}
   \put(77,32){\large $t$}
   \put(108,20){\large $b$}
\put(30,10){\psfig{file=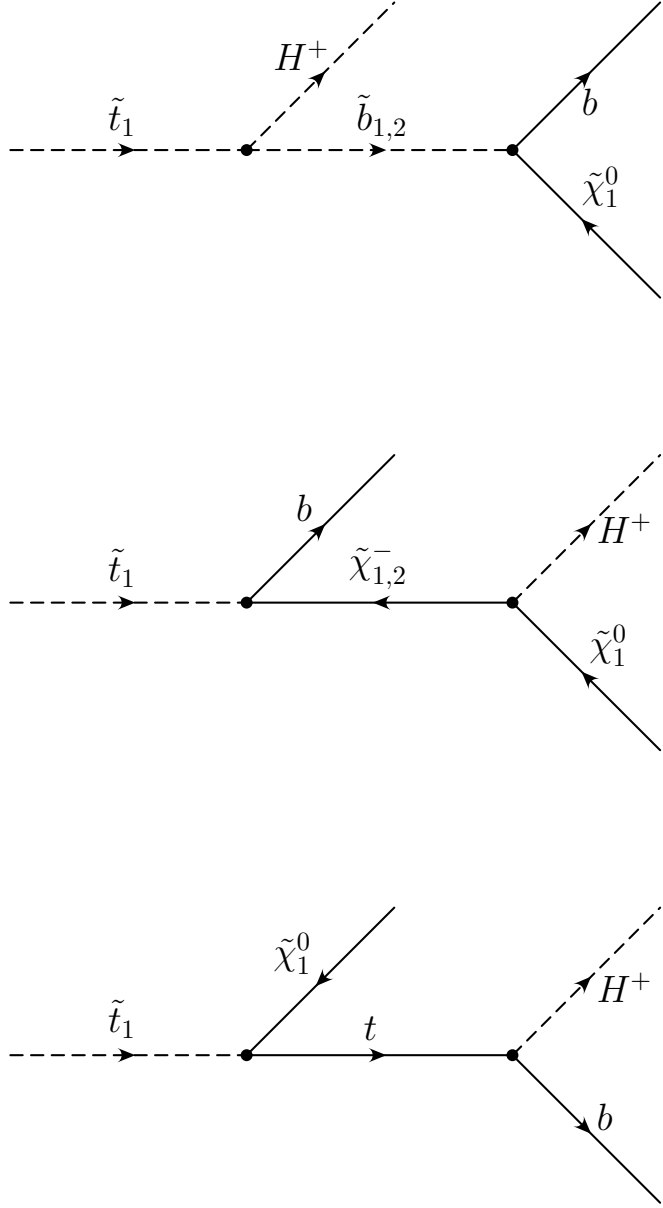,height=16cm}}
\end{picture}
\caption[]{Feynman  diagrams for the decay
    ${\tilde t}_1 \to H^+ \, b \, {\tilde \chi}^0_1$.}
\label{stbHgchi}
\end{figure}

In Fig.~\ref{stbHgchi} we show the Feynman diagrams for the decay 
${\tilde t}_1 \to H^+  b  {\tilde \chi}^0_1$.
The matrix element
$M_{\tilde t_1 \to H^+ \, b \, \tilde \chi^0_1}$
 for this decay is given by:
\begin{eqnarray}
\hspace*{-7mm}
M_{\tilde t_1 \to H^+ \, b \, \tilde \chi^0_1}
  & = &  - g^2 \sum^2_{j=1} C^H_{{\tilde t}_1 {\tilde b}_j}
     \frac{ \bar u(p_b)
           \left[ b^{\tilde{b}}_{j1} P_L + a^{\tilde{b}}_{j1} P_R \right]
           v (p_{{\tilde \chi}^0_1} )}
          {p_{{\tilde b}_j}^2 - m^2_{{\tilde b}_j}
            - i m_{{\tilde b}_j} \Gamma_{\tilde{b}_j} } \nonumber \\
 & & - \, g^2  \sum^2_{j=1}
    \frac{ \bar u(p_b)
    \left[ l^{\tilde t}_{1j} P_R + k^{\tilde t}_{1j} P_L \right]
          [\not \! p_{{\tilde \chi}^+_j} - m_{\tilde{\chi}^+_j} ]
  \left[ Q^L_{1j}{}' P_L + Q^R_{1j}{}' P_R  \right] v(p_{{\tilde \chi}^0_1} )}
  { p_{{\tilde \chi}^+_j}^2 - m^2_{\tilde{\chi}^+_j}
  - i m_{\tilde{\chi}^+_j} \Gamma_{\tilde{\chi}^+_j}} \nonumber \\
 & & + \frac{g^2} {\sqrt2 m_W }
    \frac{ \bar u (p_b )
   \left[ m_b \tan \beta P_L + m_t \cot \beta P_R \right]
        [ \not \! p_t +m_t ]
   \left[ b^{\tilde{t}}_{11} P_L + a^{\tilde{t}}_{11} P_R \right]
             v (p_{{\tilde \chi}^0_1} ) }
       {p_{t}^2 -m_{t}^2  - i m_t \Gamma_{t}} \nonumber \\
\label{eq:TfistbHchi}
\end{eqnarray}
The decay width is given by
\begin{eqnarray}
 \Gamma({\tilde t}_1 \to H^+  b  {\tilde \chi}^0_1) & = & \nonumber \\
 & & \hspace{-43mm} = \frac{\alpha^2}{16 \, \pi m^3_{{\tilde t}_1}
        \sin^4 \theta_W}
    \int\limits^{(m_{{\tilde t}_1}-m_{H^+})^2}_{
                                       (m_b + m_{{\tilde \chi}^0_1})^2}
   d \, s
   \left( G_{{\tilde \chi}^+_{} {\tilde \chi}^+_{}} +
   G_{{\tilde \chi}^+_{} t} +
   G_{{\tilde \chi}^+_{} {\tilde b}_{}} +
   G_{t t} +
   G_{t {\tilde b}_{}} +
   G_{{\tilde b}_{} {\tilde b}_{}} \right) .
\end{eqnarray}
with $G_{ij}$ given in Appendix~\ref{appA}. 

Alternatively  ${\tilde t}_1$
can decay into sleptons:
${\tilde t}_1 \to b \, \tilde{\nu}_e \, e^+,
                \,  b \, \tilde{\nu}_\mu \, \mu^+$,
${\tilde t}_1 \to b \, \tilde{e}^+_L \, \nu_{e},
         \, b \, \tilde{\mu}^+_L \, \nu_{\mu}$,
${\tilde t}_1 \to b \, {\tilde \nu}_\tau  \, \tau^+$, and
${\tilde t}_1 \to b \, {\tilde \tau}_{1,2}^+ \, \nu_{\tau}$.
These decays are mediated through virtual charginos. The Feynman graphs
are similar to the second one in Fig.~\ref{stbHgchi}, where one has
to replace the Higgs boson by a slepton and the neutralino by the corresponding
lepton.
Note, that the decays into $\tilde{e}_R$ and $\tilde{\mu}_R$ are negligible
because their couplings to the charginos are proportional to $m_e / m_W$
and $m_\mu / m_W$, respectively.
In the case of decays into sneutrinos and leptons the matrix elements
$M_{\tilde t_1 \to b \, l^+ \, \tilde \nu_l}$
have the generic form:
\begin{eqnarray}
M_{\tilde t_1 \to b \, l^+ \, \tilde \nu_l}
  =   g^2  \sum^2_{j=1} 
 \frac{ \bar u(p_b)
    \left[ l^{\tilde t}_{1j} P_R + k^{\tilde t}_{1j} P_L \right]
     \left[ \not \! p_{{\tilde \chi}^+_j} - m_{\tilde{\chi}^+_j} \right]
 \left[ l^{\tilde \nu_l}_{j} P_R + k^{\tilde \nu_l}_{j} P_L  \right] v(p_l )}
  { p_{{\tilde \chi}^+_j}^2 - m^2_{\tilde{\chi}^+_j}
  - i m_{\tilde{\chi}^+_j} \Gamma_{\tilde{\chi}^+_j}}, 
\label{eq:Tfistbsneut}
\end{eqnarray}
whereas for the decays into sleptons and neutrinos we get:
\begin{eqnarray}
M_{\tilde t_1 \to b \, \nu_l \, {\tilde l}^+_k}
 = g^2  \sum^2_{j=1} 
 \frac{ \bar u(p_b)
    \left[ l^{\tilde t}_{1j} P_R + k^{\tilde t}_{1j} P_L \right]
     \left[ \not \! p_{{\tilde \chi}^+_j} - m_{\tilde{\chi}^+_j} \right]
 \left[ l^{\tilde l}_{kj} P_L \right] v(p_{\nu_l} )}
  { p_{{\tilde \chi}^+_j}^2 - m^2_{\tilde{\chi}^+_j}
  - i m_{\tilde{\chi}^+_j} \Gamma_{\tilde{\chi}^+_j}} \, .
\label{eq:Tfistbslept}
\end{eqnarray}
In both cases the decay width is given by
\begin{eqnarray}
 \Gamma({\tilde t}_1 \to b \, \tilde{l} \, l') 
 = \frac{\alpha^2}{16 \, \pi m^3_{{\tilde t}_1}
   \sin^4 \theta_W}
 \int\limits^{(m_{{\tilde t}_1}-m_b)^2}_{(m_{l'} + m_{\tilde{l}})^2} d \, s \,
        W_{l' \tilde{l}}(s)
        \sum^3_{i=1} \left( \sum^5_{j=1} c_{ij} s^{(j-4)} \right) D_i (s) \, .
\end{eqnarray}
The explicit expressions for $W_{l' \tilde{l}}$, $c_{ij}$, and $D_i(s)$
 are given in Appendix~\ref{appA}.

\section{Numerical results}

In this section we present our numerical results for the branching ratios
of the higher order decays  of ${\tilde t}_1$. Here we consider
scenarios where all two-body decays at tree-level are kinematically forbidden.

We have fixed the parameters as in \cite{Porod97} to avoid colour breaking
minimas: we have used 
$m_{{\tilde t}_1}$, $\cos \theta_{\tilde t}$, $\tan \beta$, and $\mu$ 
as input parameters in the top squark sector.
For the sbottom (stau) sector we have fixed $M_{\tilde Q}, M_{\tilde D}$ 
and $A_b$ ($M_{\tilde E}, M_{\tilde L}$, and $A_\tau$) as input parameters. 
For simplicity, we assume that the soft SUSY breaking parameters are
equal for all generations. Note, that
due to $SU(2)$ invariance $M_{\tilde Q}$ appears in both up- and down-type
squark mass matrices.
In the sbottom (stau) sector the physical quantities 
$m_{{\tilde b}_1}$, $m_{{\tilde b}_2}$, and $\cos \theta_{\tilde b}$ 
($m_{{\tilde \tau}_1}$, $m_{{\tilde \tau}_2}$,
 and $\cos \theta_{\tilde \tau}$) obviously change with $\mu$ and $\tan \beta$.

\begin{figure}
\setlength{\unitlength}{1mm}
\begin{picture}(150,65)
\put(-4,-4){\mbox{\epsfig{figure=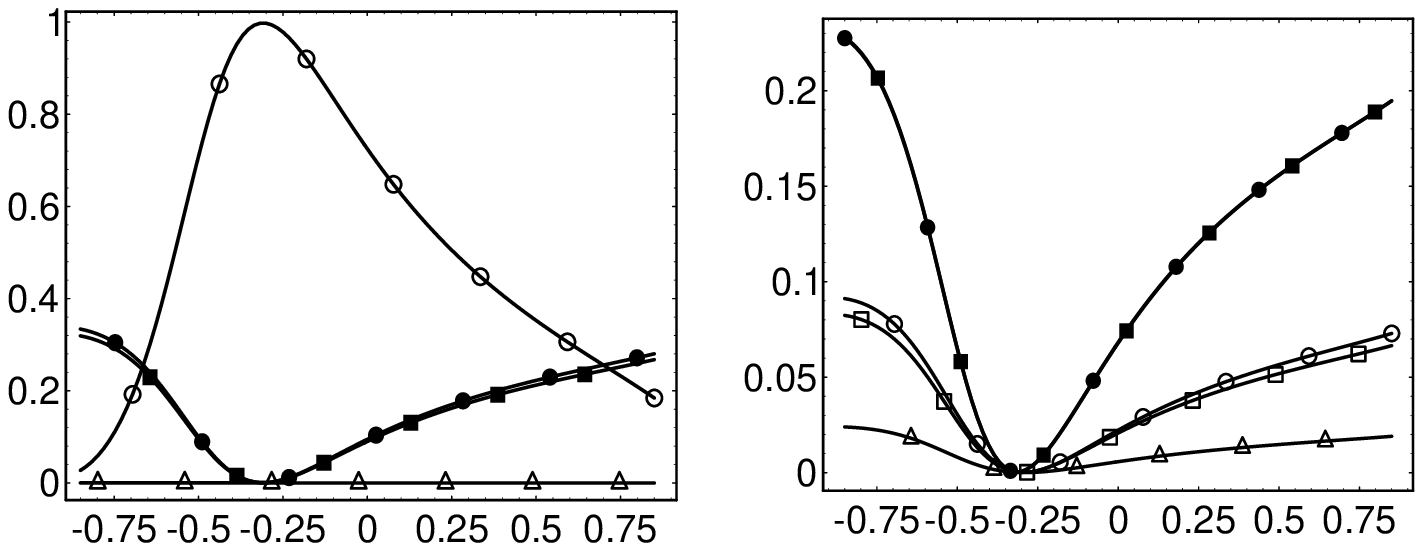,height=7.0cm,width=15.6cm}}}
\put(-1,61){{\small \bf a)}}
\put(4,60){\makebox(0,0)[bl]{{\small $\mbox{BR}({\tilde t}_1 )$}}}
\put(70.5,1){\makebox(0,0)[br]{{\small $\cos \theta_{\tilde t}$}}}
\put(75,60.5){{\small \bf b)}}
\put(80,59.5){\makebox(0,0)[bl]{{\small $\mbox{BR}({\tilde t}_1 )$}}}
\put(148.5,1.5){\makebox(0,0)[br]{{\small $\cos \theta_{\tilde t}$}}}
\end{picture}
\caption[]{Branching ratios for ${\tilde t}_1$ decays as a function of
    $\cos \theta_{\tilde t}$  for $m_{{\tilde t}_1} = 250$~GeV,
    $\tan \beta = 3$, $\mu = 530$~GeV, and
    $M = 270$~GeV. The other parameters are given in Table~I.
    The curves in a) correspond to the transitions:
           $\circ \, {\tilde t}_1 \to b \, W^+ \, {\tilde \chi}^0_1$,
           $\triangle \, {\tilde t}_1 \to c {\tilde \chi}^0_1$, 
 $\blacksquare \, ({\tilde t}_1 \to b \, e^+ \, \tilde{\nu}_e)$
 + $({\tilde t}_1 \to b \, \nu_e \, \tilde{e}^+_L)$, and
 $\bullet \, ({\tilde t}_1 \to b \, \tau^+ \, {\tilde \nu}_\tau )$
                + $({\tilde t}_1 \to b \, \nu_\tau \, {\tilde \tau}_1)$
                + $({\tilde t}_1 \to b \, \nu_\tau \, {\tilde \tau}_2)$.
           The curves in b) correspond to the transitions:
           $\circ \, {\tilde t}_1 \to b \, \nu_e \,
                    \tilde{e}^+_L$, 
 $\square \, {\tilde t}_1 \to b \, \nu_\tau \, {\tilde \tau}_1$,
 $\triangle \, {\tilde t}_1 \to b \, \nu_\tau \, {\tilde \tau}_2$,
 $\blacksquare \, {\tilde t}_1 \to b \, e^+ \, \tilde{\nu}_e$, and
 $\bullet \, {\tilde t}_1 \to b \, \tau^+ \, {\tilde \nu}_\tau $.
      }
\label{brst3cosa}
\end{figure}

In Fig.~\ref{brst3cosa}(a) and (b) we show the branching ratios of
${\tilde t}_1$ as a function of $\cos \theta_{\tilde t}$. We have restricted 
the $\cos \theta_{\tilde t}$ range such
that $|A_t| \leq 1$~TeV to avoid color/charge breaking minima. The
parameters and physical quantities are given in Tab.~\ref{tabbrst3cosa}.
The slepton parameters have been chosen such that the sum of the masses of
the final state particles are $215 \pm 5$~GeV.
In Fig.~\ref{brst3cosa}(a) we show
$\mbox{BR}({\tilde t}_1 \to b \, W^+ \, {\tilde \chi}^0_1)$,
$\mbox{BR}({\tilde t}_1 \to c \, {\tilde \chi}^0_1)$,
$\mbox{BR}({\tilde t}_1 \to b \, e^+ \, \tilde{\nu}_e$) + 
$\mbox{BR}({\tilde t}_1 \to b \, \nu_e \, \tilde{e}^+_L)$, and
$\mbox{BR}({\tilde t}_1 \to b \, \tau^+ \, {\tilde \nu}_\tau)$ +
$\mbox{BR}({\tilde t}_1 \to b \, \nu_\tau \, {\tilde \tau}^+_1)$ +
$\mbox{BR}({\tilde t}_1 \to b \, \nu_\tau \, {\tilde \tau}^+_2)$. 
Here we have not included the possibility of the decay
${\tilde t}_1 \to b \, H^+ \, {\tilde \chi}^0_1$ because
with this parameter set there exists
no value of $m_{A^0}$ which simultaneously allows this decay and fulfills the
condition $m_{h^0} \geq 71$~GeV \cite{higgsbound} 
(we have used the MSSM formula for the
calculation of $m_{H^+}$ including 1-loop corrections as given in
\cite{Brignole92}). However, we will discuss this decay later on.
We have summed up those branching ratios
for the decays into sleptons that give the same final state, for example:
\begin{eqnarray}
{\tilde t}_1 \to b \, \nu_\tau \, {\tilde \tau}^+_1 \,
          \to \, b \, \tau^+ \, \nu_\tau \, {\tilde \chi}^0_1 \;, \hspace{5mm}
{\tilde t}_1 \to b \, \tau^+ \, {\tilde \nu}_\tau \,
          \to \, b \, \tau^+ \, \nu_\tau \, {\tilde \chi}^0_1
\end{eqnarray}
Note, that in the above cases the assumption 
$m_{{\tilde t}_1} - m_b < m_{{\tilde\chi}^+_1}$
implies $m_{{\tilde\chi}^+_1} > m_{\tilde l}$. Therefore, the sleptons can 
only decay into the corresponding lepton plus
${\tilde \chi}^0_1$ except for a small parameter region where
the decay into ${\tilde \chi}^0_2$ is possible. 
However, this decay is negligible due to kinematics in that region. 
The branching ratios for decays into
$\tilde{\mu}_{L}$ or $\tilde{\nu}_{\mu}$  are 
practically the same as those into $\tilde{e}_{L}$ or $\tilde{\nu}_{e}$.
For this set of parameters $\mbox{BR}({\tilde t}_1 \to c \, {\tilde \chi}^0_1)$
 is $O(10^{-4})$ independent of $\cos \theta_{\tilde t}$ and therefore
negligible. Near $\cos \theta_{\tilde t} = -0.3$ 
$\mbox{BR}({\tilde t}_1 \to b \, W^+ \, {\tilde \chi}^0_1)$ 
is almost $100\%$ because the 
${\tilde t}_1$-${\tilde \chi}^+_1$-$b$ coupling $l^{\tilde t}_{11}$ vanishes.
We have found that the decay 
${\tilde t}_1 \to b \, W^+ \, {\tilde \chi}^0_1$ is dominated by 
the $t$-quark exchange.
In many cases the interference term between $t$ and ${\tilde \chi}^+_{1,2}$ is 
more important than
the ${\tilde \chi}^+_{1,2}$ exchange. Moreover, we have found that the 
contribution from sbottom exchange is in general negligible.

In Fig.~\ref{brst3cosa}(b) the branching ratios for the various decays into 
sleptons are shown. For small $\tan \beta$ sleptons couple mainly 
to the gaugino components of ${\tilde \chi}^+_1$. This leads to
$\mbox{BR}({\tilde t}_1 \to b \, \nu_e \, \tilde{e}^+_L) > 
\mbox{BR}({\tilde t}_1 \to b \, \nu_\tau \, \tilde{\tau}^+_{1,2})$
because $\cos \theta_{\tilde \tau} \simeq 0.68$.
The decays into sneutrinos are preferred by kinematics while the decay
into ${\tilde \tau}_2$ is suppressed by the same reason 
(Table~\ref{tabbrst3cosa}). Moreover, the
matrix elements Eqs.~(\ref{eq:Tfistbsneut}) and (\ref{eq:Tfistbslept})
for the decays into charged and neutral sleptons have
a different structure in the limit $m_b, m_l \to 0$:
\begin{eqnarray}
M_{\tilde t_1 \to b \, l^+ \, \tilde \nu_l} & \sim &  
         m_{{\tilde \chi}^+_i} \bar u(p_b) P_R v(p_l) \, ,\\
M_{\tilde t_1 \to b \, \nu_l \, {\tilde l}^+_k} & \sim &  
         \bar u(p_b) P_R  \not \! p_{{\tilde \chi}^+_i} v(p_{\nu_l}) \, .
\end{eqnarray}
This leads to different decay widths even in the limit of equal slepton
masses.

\begin{figure}
{\setlength{\unitlength}{1mm}
\begin{picture}(150,65)                        
\put(-4,-3.5){\mbox{\epsfig{figure=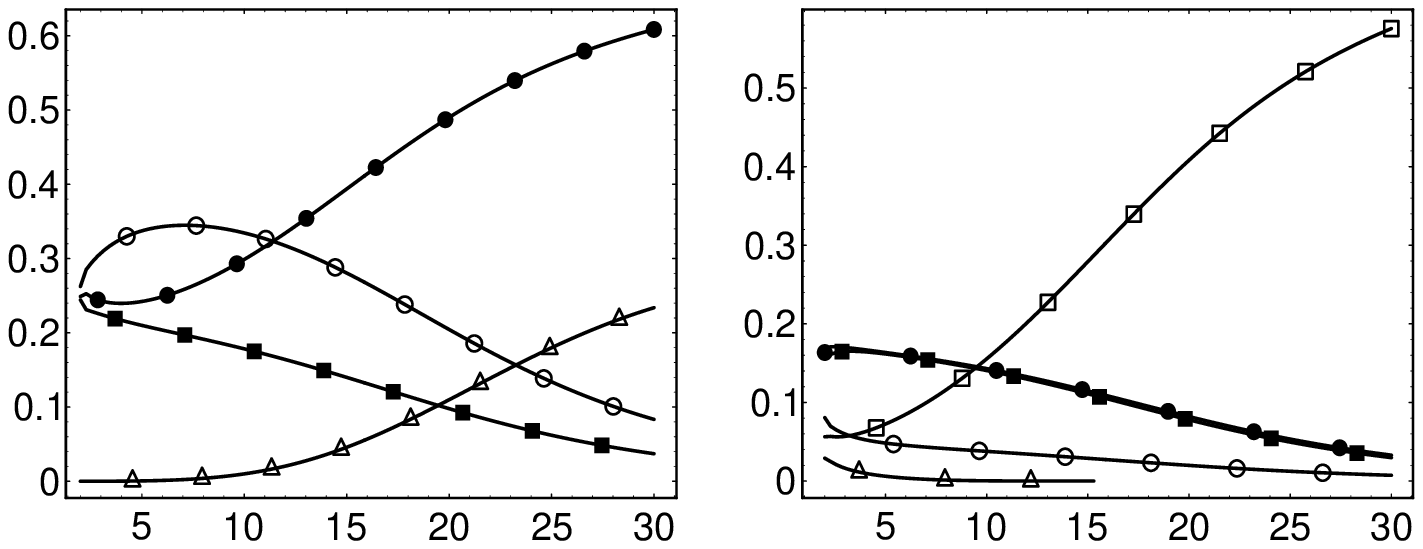,height=7.0cm,width=15.6cm}}}
\put(-1,61.5){{\small \bf a)}}
\put(4,60.5){\makebox(0,0)[bl]{{\small $\mbox{BR}({\tilde t}_1 )$}}}
\put(70.5,1){\makebox(0,0)[br]{{\small $\tan \beta$}}}
\put(75,61.5){{\small \bf b)}}
\put(80,60.5){\makebox(0,0)[bl]{{\small $\mbox{BR}({\tilde t}_1 )$}}}
\put(148.5,1.0){\makebox(0,0)[br]{{\small $\tan \beta$}}}
\end{picture}}
\caption[]{Branching ratios for ${\tilde t}_1$ decays as a function of
   $\tan \beta$   for $m_{{\tilde t}_1} = 250$~GeV,
     $\cos \theta_{\tilde t} = 0.6$, $\mu = 530$~GeV,
     $M = 270$~GeV. The other parameters are given in Table~I.
     The curves in a) correspond to the transitions:
     $\circ \, {\tilde t}_1 \to b \, W^+ \, {\tilde \chi}^0_1$,
     $\triangle \, {\tilde t}_1 \to c {\tilde \chi}^0_1$, 
 $\blacksquare \,({\tilde t}_1 \to b \, e^+ \, \tilde{\nu}_e)$
                + $({\tilde t}_1 \to b \, \nu_e \, \tilde{e}^+_L)$, and
  $\bullet \, ({\tilde t}_1 \to b \, \tau^+ \, {\tilde \nu}_\tau )$
                + $({\tilde t}_1 \to b \, \nu_\tau \, {\tilde \tau}_1)$
                + $({\tilde t}_1 \to b \, \nu_\tau \, {\tilde \tau}_2)$.
           The curves in b) correspond to the transitions:
      $\circ \, {\tilde t}_1 \to b \, \nu_e \, \tilde{e}^+_L$, 
 $\square \, {\tilde t}_1 \to b \, \nu_\tau \, {\tilde \tau}_1$,
 $\triangle \, {\tilde t}_1 \to b \, \nu_\tau \, {\tilde \tau}_2$,
 $\blacksquare \, {\tilde t}_1 \to b \, e^+ \, \tilde{\nu}_e$, and
   $\bullet \, {\tilde t}_1 \to b \, \tau^+ \, {\tilde \nu}_\tau $.
      }
\label{brst3tana}
\end{figure}

In Figs.~\ref{brst3tana}(a) and (b) we show the branching ratios as 
a function of $\tan \beta$
for $\cos \theta_{\tilde t} = 0.6$ and the other parameters as above. For small
$\tan \beta$ the
decay into ${\tilde t}_1 \to b \, W^+ \, {\tilde \chi}^0_1$ is the most 
important one. The branching ratios for the decays into sleptons decrease 
with increasing $\tan \beta$ except for the decay into ${\tilde \tau}_1$.
This results from: (i) for increasing $\tan \beta$ the
gaugino component of ${\tilde \chi}^+_1$ decreases while its mass increases,
(ii) the masses of the sleptons increase with increasing $\tan \beta$, except
$m_{{\tilde \tau}_1}$ which decreases, and (iii) the $\tau$ Yukawa coupling
increases. These facts lead to the dominance of 
${\tilde t}_1 \to b \, \nu_\tau \, {\tilde \tau}_1$ for large 
$\tan \beta$ as can be seen in Fig.~\ref{brst3tana}(b). In addition, 
the decay into $c \, {\tilde \chi}^0_1$ gains some importance for large
$\tan \beta$ because its width is proportional to the bottom
Yukawa coupling in the approximation of \cite{Hikasa87}.

\begin{figure}[t]
{\setlength{\unitlength}{1mm}
\begin{picture}(150,65)                        
\put(-4,-3.5){\mbox{\epsfig{figure=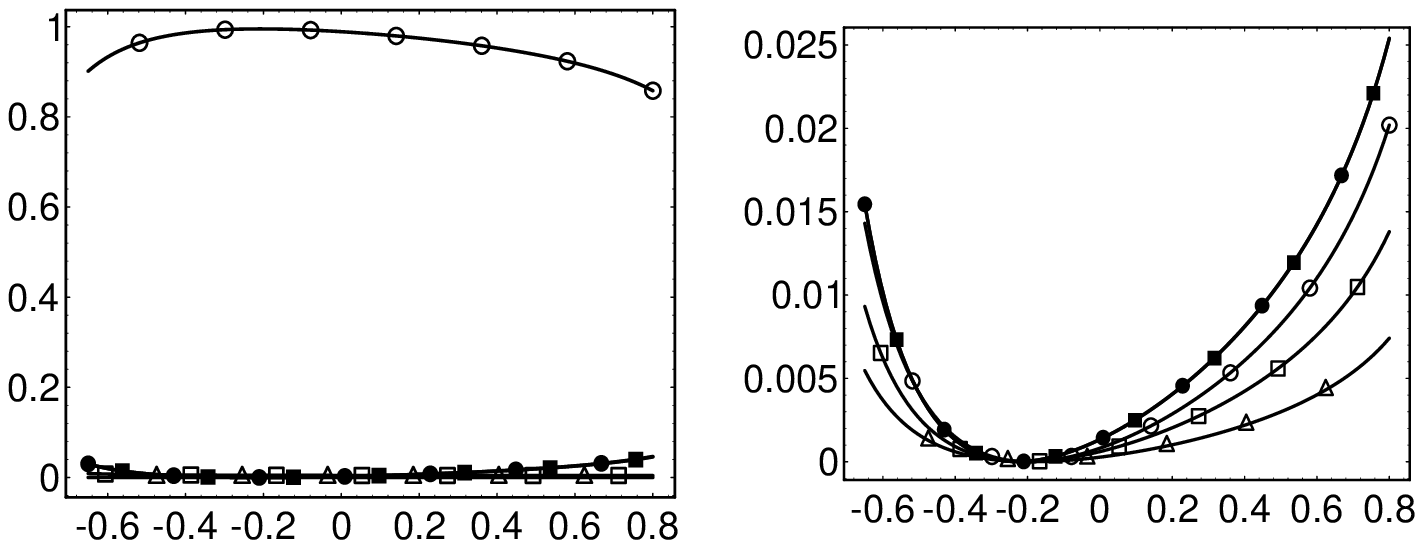,height=7.0cm,width=15.6cm}}}
\put(-1,61){{\small \bf a)}}
\put(4,60){\makebox(0,0)[bl]{{\small $\mbox{BR}({\tilde t}_1 )$}}}
\put(70.5,1){\makebox(0,0)[br]{{\small $\cos \theta_{\tilde t}$}}}
\put(75,60){{\small \bf b)}}
\put(80,59){\makebox(0,0)[bl]{{\small $\mbox{BR}({\tilde t}_1 )$}}}
\put(148.5,2.0){\makebox(0,0)[br]{{\small $\cos \theta_{\tilde t}$}}}
\end{picture}}
\caption[]{Branching ratios for ${\tilde t}_1$ decays as a
  function of $\cos \theta_{\tilde t}$
    for $m_{{\tilde t}_1} = 350$~GeV, $\tan \beta = 3$, $\mu = 750$~GeV,
    $M = 380$~GeV, and $m_{A^0} = 110$~GeV. The other parameters are given 
     in Table~II.  The curves in a) correspond to the transitions:
   $\circ \, {\tilde t}_1 \to b \, W^+ \, {\tilde \chi}^0_1$,
 $\square \, {\tilde t}_1 \to b \,H^+ \,  {\tilde \chi}^0_1$, 
           $\triangle \, {\tilde t}_1 \to c {\tilde \chi}^0_1$, 
 $\blacksquare \, ({\tilde t}_1 \to b \, e^+ \, \tilde{\nu}_e)$
           + $({\tilde t}_1 \to b \, \nu_e \, \tilde{e}^+_L)$, and
 $\bullet \, ({\tilde t}_1 \to b \, \tau^+ \, {\tilde \nu}_\tau )$
                + $({\tilde t}_1 \to b \, \nu_\tau \, {\tilde \tau}_1)$
                + $({\tilde t}_1 \to b \, \nu_\tau \, {\tilde \tau}_2)$.
           The curves in b) correspond to the transitions:
    $\circ \, {\tilde t}_1 \to b \, \nu_e \, \tilde{e}^+_L$, 
 $\square \, {\tilde t}_1 \to b \, \nu_\tau \, {\tilde \tau}_1$,
  $\triangle \, {\tilde t}_1 \to b \, \nu_\tau \, {\tilde \tau}_2$,
 $\blacksquare \, {\tilde t}_1 \to b \, e^+ \, \tilde{\nu}_e$, and
  $\bullet \, {\tilde t}_1 \to b \, \tau^+ \, {\tilde \nu}_\tau $.
      }
\label{brst3cospmu}
\end{figure}

The assumption that no two-body decays be allowed at
tree level implies that $m_{{\tilde\chi}^+_1} > m_{{\tilde t}_1} - m_b$. 
Therefore, one expects
an increase of $\mbox{BR}({\tilde t}_1 \to b \, W^+ \, {\tilde \chi}^0_1)$ 
if $m_{{\tilde t}_1}$ increases, because
the decay into $b \, W^+ \, {\tilde \chi}^0_1$ is dominated by the $t$ exchange
whereas for the decays into sleptons ${\tilde \chi}^+_1$ exchange dominates. 
 This is demonstrated in Figs.~\ref{brst3cospmu}(a) and (b) where we
have fixed $m_{{\tilde t}_1} = 350$~GeV. 
Here also the decay into $b \, H^+ \, {\tilde \chi}^0_1$ is possible. However,
this channel is in general suppressed by kinematics. We have not
found any case with $m_{H^+} \leq 120$~GeV while $m_{h^0} > 75$~GeV
\cite{higgsbound}.

\begin{figure}[t]
{\setlength{\unitlength}{1mm}
\begin{picture}(150,65)                        
\put(-4,-5){\mbox{\epsfig{figure=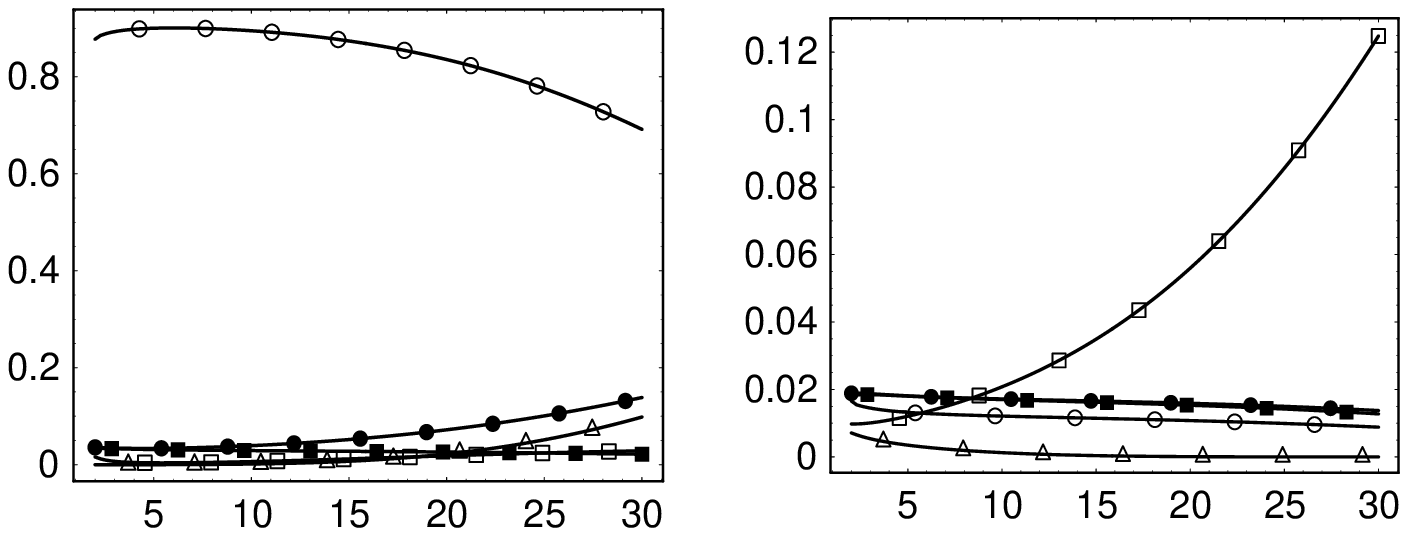,height=7.0cm,width=15.6cm}}}
\put(-1,60.5){{\small \bf a)}}
\put(4,59.5){\makebox(0,0)[bl]{{\small $\mbox{BR}({\tilde t}_1 )$}}}
\put(70.5,1){\makebox(0,0)[br]{{\small $\tan \beta$}}}
\put(75,59.5){{\small \bf b)}}
\put(80,58.5){\makebox(0,0)[bl]{{\small $\mbox{BR}({\tilde t}_1 )$}}}
\put(148.5,2.0){\makebox(0,0)[br]{{\small $\tan \beta$}}}
\end{picture}}
\caption[]{Branching ratios for ${\tilde t}_1$ decays as a function of
       $\tan \beta$  for $m_{{\tilde t}_1} = 350$~GeV, 
       $\cos \theta_{\tilde t} = 0.7$, $\mu = 750$~GeV,
        $M = 380$~GeV and $m_{A^0} = 110$~GeV. The other parameters are given 
      in Table~II. The curves in a) correspond to the transitions:
     $\circ \, {\tilde t}_1 \to b \, W^+ \, {\tilde \chi}^0_1$,
$\square \, {\tilde t}_1 \to b \,H^+ \,  {\tilde \chi}^0_1$, 
           $\triangle \, {\tilde t}_1 \to c {\tilde \chi}^0_1$, 
$\blacksquare \, ({\tilde t}_1 \to b \, e^+ \, \tilde{\nu}_e)$
                + $({\tilde t}_1 \to b \, \nu_e \, \tilde{e}^+_L)$, and
  $\bullet \, ({\tilde t}_1 \to b \, \tau^+ \, {\tilde \nu}_\tau )$
                + $({\tilde t}_1 \to b \, \nu_\tau \, {\tilde \tau}_1)$
                + $({\tilde t}_1 \to b \, \nu_\tau \, {\tilde \tau}_2)$.
           The curves in b) correspond to the transitions:
     $\circ \, {\tilde t}_1 \to b \, \nu_e \, \tilde{e}^+_L$, 
 $\square \, {\tilde t}_1 \to b \, \nu_\tau \, {\tilde \tau}_1$,
 $\triangle \, {\tilde t}_1 \to b \, \nu_\tau \, {\tilde \tau}_2$,
 $\blacksquare \, {\tilde t}_1 \to b \, e^+ \, \tilde{\nu}_e$, and
  $\bullet \, {\tilde t}_1 \to b \, \tau^+ \, {\tilde \nu}_\tau $.
      }
\label{brst3tanb}
\end{figure}

\begin{figure}
\begin{center}
{\setlength{\unitlength}{1mm}
\begin{picture}(77,72)
\put(-0.1,2.0){\mbox{\psfig{figure=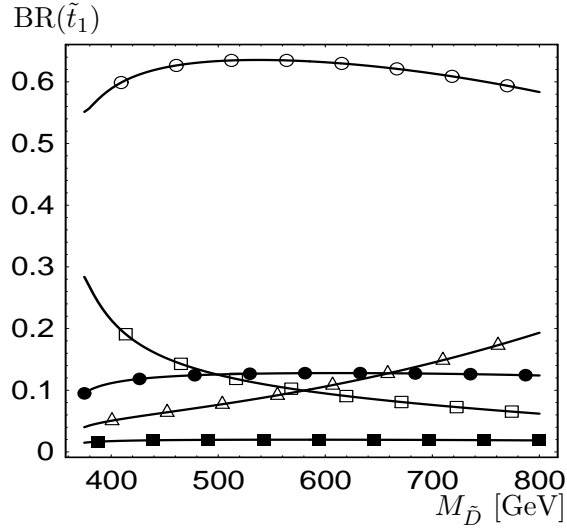,height=6.3cm,width=7.5cm}}}
\put(1,66){\makebox(0,0)[bl]{{\small $\mbox{BR}({\tilde t}_1 )$}}}
\put(74.5,1){\makebox(0,0)[br]{{\small $M_{\tilde D}$~[GeV]}}}
\end{picture}}
\end{center}
\caption[]{Branching ratios for ${\tilde t}_1$ decays as a function of 
      $M_{\tilde D}$
            for $m_{{\tilde t}_1} = 350$~GeV, $\cos \theta_{\tilde t} = 0.7$,
            $\tan \beta = 30$,  $\mu = 750$~GeV,  and
            $m_{A^0} = 90$~GeV. The other input parameters are the same as 
           in Table~II.
           The curves in a) correspond to the transitions:
    $\circ \, {\tilde t}_1 \to b \, W^+ \, {\tilde \chi}^0_1$,
 $\square \, {\tilde t}_1 \to b \,H^+ \,  {\tilde \chi}^0_1$, 
           $\triangle \, {\tilde t}_1 \to c {\tilde \chi}^0_1$, 
 $\blacksquare \, ({\tilde t}_1 \to b \, e^+ \, \tilde{\nu}_e)$ 
                + $({\tilde t}_1 \to b \, \nu_e \, \tilde{e}^+_L)$, and
 $\bullet \, ({\tilde t}_1 \to b \, \tau^+ \, {\tilde \nu}_\tau )$
                + $({\tilde t}_1 \to b \, \nu_\tau \, {\tilde \tau}_1)$
                + $({\tilde t}_1 \to b \, \nu_\tau \, {\tilde \tau}_2)$.
      }
\label{brst3MD}
\end{figure}

These general features still hold if $\tan \beta$ increases as can be seen in
Figs.~\ref{brst3tanb}(a) and (b). Here we have fixed 
$\cos \theta_{\tilde t} = 0.7$.
In accordance with the discussion above, the decay into 
${\tilde t}_1 \to b \, \nu_\tau \, {\tilde \tau}_1$ gains importance 
with increasing $\tan \beta$. Note, that for large $\tan \beta$
$\mbox{BR}({\tilde t}_1 \to b \, H^+ \, {\tilde \chi}^0_1)$ 
decreases since $m_{H^+}$ increases due
to radiative corrections. However, there are scenarios where the decay
${\tilde t}_1 \to b \, H^+ \, {\tilde \chi}^0_1$ 
becomes important. This can be seen in Fig.~\ref{brst3MD} where
the branching ratios are shown as a function of $M_{\tilde D}$ 
for $m_{A^0} = 90$~GeV,
$\tan \beta = 30$, and the other parameters as in Tab.~\ref{tabbrst3cospmu}.
At the lower end of the $M_{\tilde D}$ range we get $m_{H^+} = 114$~GeV. 
Moreover, $m_{{\tilde b}_1}$ is approximately $m_{{\tilde t}_1} - m_W$ 
leading to an enhancement of this width. We have found that contrary
to the case ${\tilde t}_1 \to b \, W^+ \, {\tilde \chi}^0_1$ for the decays
${\tilde t}_1 \to b \, H^+ \, {\tilde \chi}^0_1$ sbottom exchange can be 
important.
This is a consequence of the different spin
structure of the corresponding matrix elements 
(Eqs.~(\ref{eq:TfistbWchi}) and
(\ref{eq:TfistbHchi})) and because of the large bottom Yukawa coupling. 
The decrease in  
$\mbox{BR}({\tilde t}_1 \to b \, H^+ \, {\tilde \chi}^0_1)$ for
$M_{\tilde D} \geq 450$~GeV is mainly due to the fact that 
$m_{H^+}$ grows with increasing $M_{\tilde D}$.

Finally, we want to discuss a scenario which is within the reach of 
an upgraded Tevatron. Here we refer to the examples of \cite{Porod97}.
In general the decays into sleptons clearly dominate
when they are kinematically allowed (except the case when the couplings of 
the top squark
to the lighter chargino nearly vanishes). In the case
$m_{{\tilde t}_1} = 170$~GeV, $\cos \theta_{\tilde t} = -0.7$, 
$M_{\tilde D} = M_{\tilde Q} = 500$~GeV, $A_b = A_\tau = -350$~GeV,
$\mu = -1000$~GeV, $M = 165$~GeV and $\tan \beta =2$ 
(scenario b of Table~I in \cite{Porod97}) we obtain:
$\mbox{BR}({\tilde t}_1 \to b \, \nu_e \, \tilde{e}^+_L) = 2.8 \%$, 
$\mbox{BR}({\tilde t}_1 \to b \, \nu_\tau \, {\tilde \tau}_1) = 10.5 \%$,
$\mbox{BR}({\tilde t}_1 \to b \, \nu_\tau \, {\tilde \tau}_2) = 
2 \times \, 10^{-3} \, \%$,
$\mbox{BR}({\tilde t}_1 \to b \, e^+ \, \tilde{\nu}_e) = 28.1 \%$, and
$\mbox{BR}({\tilde t}_1 \to b \, \tau^+ \, {\tilde \nu}_\tau) = 27.8 \%$. 
The order
of magnitude is independent of $\mu$ and $\cos \theta_{\tilde t}$ because
the lighter chargino is mainly gaugino-like in the parameter space where 
the decay ${\tilde t}_1 \to b \, W^+ \, {\tilde \chi}^0_1$ is possible
(see Fig.~2b of \cite{Porod97}).
For increasing $\tan \beta$ we have found a similar behaviour as in
Figs.~\ref{brst3tana} and ~\ref{brst3tanb}:
dominance of the decay ${\tilde t}_1 \to b \, \nu_\tau \, {\tilde \tau}^+_1$,
an increase of ${\tilde t}_1 \to c \, {\tilde \chi}^0_1$ and a decrease of
all other decay channels.

\section{Conclusions}

We have calculated the three-body decays 
${\tilde t}_1 \to b \, H^+ \, {\tilde \chi}^0_1$,
${\tilde t}_1 \to b \, \nu_l \, {\tilde l}^+_i$, and
${\tilde t}_1 \to b \, l^+ \, {\tilde \nu}_l$ ($l=e,\mu,\tau$) including
all terms proportional to
$m_b$, $m_\tau$, and all Yukawa couplings. We have compared these decays
with  ${\tilde t}_1 \to c \, {\tilde \chi}^0_1$ \cite{Hikasa87} and
${\tilde t}_1 \to b \, W^+ \, {\tilde \chi}^0_1$ \cite{Porod97}.
These decays are competitive in that part of the parameter space --- accessible
at an upgraded Tevatron, the LHC or a future lepton collider --- where
the tree--level two-body decays ${\tilde t}_1 \to b \, {\tilde \chi}^+_i$ and
${\tilde t}_1 \to t \, {\tilde \chi}^0_j$ are kinematically forbidden. 
We have found that for $m_{{\tilde t}_1}\leq 200$ GeV the decays
into sleptons dominate due to kinematics. 
In the range 200 GeV $\leq m_{{\tilde t}_1} \leq 300$ GeV all the 
decays mentioned  above compete with each other.
The branching ratios depend
crucially on the coupling of ${\tilde t}_1$ to ${\tilde \chi}^+_1$,
implying that one can get information
on the mixing angle of the top squarks, once the chargino properties are known.
For heavier top squark masses the decay into  
$b \, W^+ \, {\tilde \chi}^0_1$, mainly proceeding via a virtual $t$ quark,
is in general the most important one.

In addition, we
have found that for small $\tan \beta$ the decays into sneutrinos are
more important than the decays into charged sleptons. This is a result of the
different spin structures of the corresponding matrix elements. For large
$\tan \beta$ the decay into the lighter stau becomes important due to
the large tau Yukawa coupling (implying also a smaller stau mass).

The decay into $b \, H^+ \, {\tilde \chi}^0_1$
is kinematically suppressed because the existing mass bounds on
the neutral Higgs bosons also imply a lower bound on $m_{H^+}$.
However, in scenarios where radiative corrections decrease $m_{H^+}$ and
where at the same time $m_{{\tilde t}_1}> m_{{\tilde b}_1}$ we have found
branching ratios of the order of 30\%.  

The large variety of possible three-body decay modes implies the chance to
determine the properties of ${\tilde t}_1$ also when higher order decays
are dominant. Clearly, 
a detailed Monte Carlo study will be necessary to see how  the different
channels can be separated.

\section*{Acknowledgments}

I thank A.~Bartl, H.~Eberl, T.~Gajdosik, S.~Kraml, and W.~Majerotto for many
helpful discussions and the inspiring atmosphere. 
I am grateful to J.W.F.~Valle for the kind hospitality
and the pleasant atmosphere at the 
Departamento de F\'{\i}sica Te\'orico.
This work was supported by the ``Fonds zur F\"orderung der wissenschaftlichen
Forschung'' of Austria, project No. P10843-PHY, and by the EEC 
under the TMR contract ERBFMRX-CT96-0090.

\begin{appendix}

\section{Formulas for the three-body decay widths}
\label{appA}

In the subsequent sections the formulas for the decay widths  are listed 
which have been omitted in Sec.~\ref{sec:threebody}.
 
\subsection{The width $\Gamma({\tilde t}_1 \to H^+ \, b \, {\tilde \chi}^0_1)$}

\noindent
The decay width is given by
\begin{eqnarray}
 \Gamma({\tilde t}_1 \to H^+ \, b \, {\tilde \chi}^0_1)  &=& \nonumber \\
 & & \hspace{-30mm} 
   =  \frac{\alpha^2}{16 \, \pi m^3_{{\tilde t}_1} \sin^4 \theta_W}
  \int\limits^{(m_{{\tilde t}_1}-m_{H^+})^2}_{
           (m_b + m_{{\tilde \chi}^0_1})^2} \hspace{-8mm}
     d \, s \,
   \left( G_{{\tilde \chi}^+ {\tilde \chi}^+} +
   G_{{\tilde \chi}^+ t} +
   G_{{\tilde \chi}^+ {\tilde b}} +
   G_{t t} +
   G_{t {\tilde b}} +
   G_{{\tilde b} {\tilde b}} \right) 
\end{eqnarray}
with
\begin{eqnarray} 
   G_{{\tilde \chi}^+ {\tilde \chi}^+} &=&
       \sum^2_{i=1} \Big[  ( a_{i1} + a_{i2} s )
          J^0_t(m^2_{{\tilde t}_1} + m^2_{H^+} + m^2_b
               + m^2_{{\tilde \chi}^0_1} - m^2_{{\tilde \chi}^+_i} - s
             ,\Gamma_{{\tilde \chi}^+_i} m_{{\tilde \chi}^+_i}) \nonumber \\
    & & \hspace{5mm} + ( a_{i3} + a_{i4} s )
       J^1_t(m^2_{{\tilde t}_1} + m^2_{H^+} + m^2_b
            + m^2_{{\tilde \chi}^0_1} - m^2_{{\tilde \chi}^+_i} - s
             ,\Gamma_{{\tilde \chi}^+_i} m_{{\tilde \chi}^+_i}) \nonumber \\
    & & \hspace{5mm} + \, a_{i4} \,
          J^2_t(m^2_{{\tilde t}_1} + m^2_{H^+} + m^2_b
               + m^2_{{\tilde \chi}^0_1} - m^2_{{\tilde \chi}^+_i} - s
       ,\Gamma_{{\tilde \chi}^+_i} m_{{\tilde \chi}^+_i}) \Big] \nonumber \\
    & &  +  ( a_{31} + a_{32} s )
          J^0_{tt}(m^2_{{\tilde t}_1} + m^2_{H^+} + m^2_b
               + m^2_{{\tilde \chi}^0_1} - m^2_{{\tilde \chi}^+_1} - s
             ,\Gamma_{{\tilde \chi}^+_1} m_{{\tilde \chi}^+_1}  \nonumber \\
       & & \hspace{33mm} ,m^2_{{\tilde t}_1} + m^2_{H^+} + m^2_b
               + m^2_{{\tilde \chi}^0_1} - m^2_{{\tilde \chi}^+_2} - s
             ,\Gamma_{{\tilde \chi}^+_2} m_{{\tilde \chi}^+_2}) \nonumber \\
    & &  + ( a_{33} + a_{34} s )
       J^1_{tt}(m^2_{{\tilde t}_1} + m^2_{H^+} + m^2_b
               + m^2_{{\tilde \chi}^0_1} - m^2_{{\tilde \chi}^+_1} - s
             ,\Gamma_{{\tilde \chi}^+_1} m_{{\tilde \chi}^+_1}  \nonumber \\
       & & \hspace{33mm} ,m^2_{{\tilde t}_1} + m^2_{H^+} + m^2_b
               + m^2_{{\tilde \chi}^0_1} - m^2_{{\tilde \chi}^+_2} - s
             ,\Gamma_{{\tilde \chi}^+_2} m_{{\tilde \chi}^+_2}) \nonumber \\
    & &  + \, a_{34} \,
       J^2_{tt}(m^2_{{\tilde t}_1} + m^2_{H^+} + m^2_b
               + m^2_{{\tilde \chi}^0_1} - m^2_{{\tilde \chi}^+_1} - s
             ,\Gamma_{{\tilde \chi}^+_1} m_{{\tilde \chi}^+_1}  \nonumber \\
       & & \hspace{17mm} ,m^2_{{\tilde t}_1} + m^2_{H^+} + m^2_b
               + m^2_{{\tilde \chi}^0_1} - m^2_{{\tilde \chi}^+_2} - s
             ,\Gamma_{{\tilde \chi}^+_2} m_{{\tilde \chi}^+_2}) \, ,
\end{eqnarray}
\begin{eqnarray}
   G_{{\tilde \chi}^+ t} &=&
       \sum^2_{i=1} \Big[ ( b_{i1} + b_{i2} s ) 
 J^0_{tt}(m^2_{{\tilde t}_1} + m^2_{H^+} + m^2_b
               + m^2_{{\tilde \chi}^0_1} - m^2_{{\tilde \chi}^+_1} - s
       ,-\Gamma_{{\tilde \chi}^+_1} m_{{\tilde \chi}^+_1},m^2_t,
        \Gamma_t m_t) \nonumber \\
    & & \hspace{5mm} + ( b_{i3} + b_{i4} s )
 J^1_{tt}(m^2_{{\tilde t}_1} + m^2_{H^+} + m^2_b
               + m^2_{{\tilde \chi}^0_1} - m^2_{{\tilde \chi}^+_1} - s
       ,-\Gamma_{{\tilde \chi}^+_1} m_{{\tilde \chi}^+_1} ,m^2_t,
        \Gamma_t m_t) \nonumber \\
    & & \hspace{5mm} + \, b_{i4} \,
        J^2_{tt}(m^2_{{\tilde t}_1} + m^2_{H^+} + m^2_b
               + m^2_{{\tilde \chi}^0_1} - m^2_{{\tilde \chi}^+_1} - s
       ,-\Gamma_{{\tilde \chi}^+_1} m_{{\tilde \chi}^+_1} ,m^2_t,
         \Gamma_t m_t)  \Big] \, ,
\end{eqnarray}
\begin{eqnarray}
   G_{{\tilde \chi}^+ {\tilde b}} &=&
       \sum^2_{k,i=1} \Big[
          ( c_{ik1} + c_{ik2} s ) \nonumber \\
    & & \hspace{7mm} * J^0_{st}(
       m^2_{{\tilde b}_k}, \Gamma_ {{\tilde b}_k} m_{{\tilde b}_k}
       ,m^2_{{\tilde t}_1} + m^2_{H^+} + m^2_b
               + m^2_{{\tilde \chi}^0_1} - m^2_{{\tilde \chi}^+_1} - s
       ,-\Gamma_{{\tilde \chi}^+_1} m_{{\tilde \chi}^+_1}) \nonumber \\
    & &  + \, c_{ik3} \, J^1_{st}(
       m^2_{{\tilde b}_k}, \Gamma_ {{\tilde b}_k} m_{{\tilde b}_k}
       ,m^2_{{\tilde t}_1} + m^2_{H^+} + m^2_b
               + m^2_{{\tilde \chi}^0_1} - m^2_{{\tilde \chi}^+_1} - s
       ,-\Gamma_{{\tilde \chi}^+_1} m_{{\tilde \chi}^+_1}) \Big] \, ,
\end{eqnarray}
\begin{eqnarray}
   G_{t t} &=&
          ( d_1 + d_2 s) J^0_t(m^2_t, \Gamma_t m_t) +
          ( d_{3} + d_{4} s) J^1_t(m^2_t, \Gamma_t m_t)
 + \, d_{4} \, J^2_t(m^2_t, \Gamma_t m_t) \, ,
\end{eqnarray}
\begin{eqnarray}
   G_{t {\tilde b}} &=& \sum^2_{k=1} \Big[
   ( e_{k1} + e_{k2} s )
   J^0_{st}(m^2_{{\tilde b}_k},
  \Gamma_ {{\tilde b}_k} m_{{\tilde b}_k},m^2_t, \Gamma_t m_t)
 + \, e_{k3} \, J^1_{st}(m^2_{{\tilde b}_k},
    \Gamma_ {{\tilde b}_k} m_{{\tilde b}_k},m^2_t, \Gamma_t m_t)
   \Big] \, , \nonumber \\
\end{eqnarray}
\begin{eqnarray}
G_{{\tilde b} {\tilde b}} &=&
 \frac{\sqrt{\lambda(s,m^2_{{\tilde t}_1},m^2_{H^+})
                 \lambda(s,m^2_{{\tilde \chi}^0_1},m^2_b)}}{s} \nonumber \\
  & & \hspace{-12mm} * \left\{ \sum^2_{k=1}
      \frac{(f_{k1} + f_{k2} s)}
   {(s-m^2_{{\tilde b}_k})^2 + \Gamma^2_{{\tilde b}_k} m^2_{{\tilde b}_k}}
 + \mbox{Re} \left[ \frac{(f_{31} + f_{32} s)}
      { (s-m^2_{{\tilde b}_1} + i \Gamma_{{\tilde b}_1} m_{{\tilde b}_1})
         (s-m^2_{{\tilde b}_2} - i \Gamma_{{\tilde b}_2} m_{{\tilde b}_2})}
          \right]  \right\} \, .
\end{eqnarray}
The integrals $J^{0,1,2}_{t,tt,st}$ are:
\begin{eqnarray}
 J^i_{t}(m^2_1, m_1 \Gamma_1)
        &=& \int\limits^{t_{max}}_{t_{min}}
     \hspace{-2mm} d \, t \frac{t^i}{(t-m^2_1)^2 \, + \,  m_1^2 \Gamma_1^2}
  \, , \\
 J^i_{tt}(m^2_1,m_1 \, \Gamma_1,m^2_2,m_2 \, \Gamma_2)
        &=& \mbox{Re} \int\limits^{t_{max}}_{t_{min}}
     \hspace{-2mm} d \, t
          \frac{t^i}{(t-m^2_1 \, + \,i m_1 \Gamma_1)
                     (t-m^2_2 \, - \,i m_2 \Gamma_2)} \, , \\
 J^i_{st}(m^2_1,m_1 \, \Gamma_1,m^2_2,m_2 \, \Gamma_2)
        &=& \mbox{Re} \frac1{s-m^2_1 \, + \,i  m_1 \Gamma_1}
         \int\limits^{t_{max}}_{t_{min}} \hspace{-2mm} d \, t
          \frac{t^i}{(t-m^2_2 \, - \,i m_2 \Gamma_2)} 
\end{eqnarray}
with $i=0,1,2$. Their integration range is given by
\begin{eqnarray}
t_{max \atop min} &=&
 \frac{m^2_{{\tilde t}_1} + m^2_b + m^2_{H^+}
        + m^2_{{\tilde \chi}^0_1} -s}2
       - \frac{(m^2_{{\tilde t}_1}-m^2_{H^+})
               (m^2_{{\tilde \chi}^0_1}-m^2_b)}{2 s} \nonumber \\
  & &  \pm  \frac{\sqrt{\lambda(s,m^2_{{\tilde t}_1},m^2_{H^+})
                  \lambda(s,m^2_{{\tilde \chi}^0_1},m^2_b)}}{2 s} \, ,
\end{eqnarray}
where $s = (p_{{\tilde t}_1} - p_{H^+})^2$
 and $t = (p_{{\tilde t}_1} - p_{t})^2$
are the usual Mandelstam variables.
Note, that $-\Gamma_{{\tilde \chi}^+_1} m_{{\tilde \chi}^+_1}$
appears in the entries of the integrals 
$G_{{\tilde \chi}^+ {\tilde b}_j}$ and
$G_{{\tilde \chi}^+ t}$ because the chargino is exchanged 
 in the $u$-channel in our convention.
The coefficients are given by:
\begin{eqnarray}
a_{11} &=& - 4 \, k^{\tilde t}_{11} l^{\tilde t}_{11} 
       Q^L_{11}{}' Q^R_{11}{}' m_b m_{{\tilde \chi}^0_1}
    \left( m^2_b +  m^2_{\tilde{\chi}^+_1}
   + m^2_{{\tilde \chi}^0_1} + m^2_{{\tilde t}_1} 
   + m^2_{H^+} \right) \nonumber \\
  & &  - 2 \, Q^L_{11}{}' Q^R_{11}{}'
    \left( (k^{\tilde t}_{11})^2 + (l^{\tilde t}_{11})^2 \right)
    m_{{\tilde \chi}^0_1} m_{\tilde{\chi}^+_1}
    \left( 2 \, m^2_b + m^2_{{\tilde \chi}^0_1}
      + m^2_{H^+} \right)  \nonumber \\
  & &  - 2 \, k^{\tilde t}_{11} l^{\tilde t}_{11} 
        \left( (Q^L_{11}{}')^2 + (Q^R_{11}{}')^2 \right)
     m_b m_{\tilde{\chi}^+_1} 
        \left( m^2_b + 2 \, m^2_{{\tilde \chi}^0_1}
            + m^2_{{\tilde t}_1} \right)  \nonumber \\
  & &  - \left( (k^{\tilde t}_{11})^2
  (Q^R_{11}{}')^2 + (l^{\tilde t}_{11})^2 (Q^L_{11}{}')^2 \right) \nonumber \\
   & & \hspace{5mm} * \left[ \left( m^2_b+m^2_{{\tilde \chi}^0_1} \right)^2
    + \left( m^2_b+m^2_{H^+} \right)
 \left(m^2_{{\tilde \chi}^0_1}+m^2_{{\tilde t}_1}\right) \right] \nonumber \\
  & &  - \left( (k^{\tilde t}_{11})^2
       (Q^L_{11}{}')^2 + (l^{\tilde t}_{11})^2 (Q^R_{11}{}')^2 \right)
      m^2_{\tilde{\chi}^+_1}
           \left(m^2_b+m^2_{{\tilde \chi}^0_1} \right) \, ,\\
a_{12} &=& 4 \, k^{\tilde t}_{11} l^{\tilde t}_{11}
       Q^L_{11}{}' Q^R_{11}{}' m_b m_{{\tilde \chi}^0_1}
  + \left( (k^{\tilde t}_{11})^2 
       (Q^R_{11}{}')^2 + (l^{\tilde t}_{11})^2 (Q^L_{11}{}')^2 \right)
           \left( m^2_b+m^2_{{\tilde \chi}^0_1} \right) \nonumber \\
  & & + 2 \, k^{\tilde t}_{11} 
       l^{\tilde t}_{11} \left( (Q^L_{11}{}')^2 + (Q^R_{11}{}')^2 \right)
                 m_b m_{\tilde{\chi}^+_1}
        + 2 \, Q^L_{11}{}' Q^R_{11}{}'
        \left( (k^{\tilde t}_{11})^2 + (l^{\tilde t}_{11})^2 \right)
         m_{{\tilde \chi}^0_1} m_{\tilde{\chi}^+_1}  \nonumber \\
  & &  + \left( (k^{\tilde t}_{11})^2
      (Q^L_{11}{}')^2 + (l^{\tilde t}_{11})^2 (Q^R_{11}{}')^2 \right)
              m^2_{\tilde{\chi}^+_1}   \, , \\
a_{13} &=& 4 \, k^{\tilde t}_{11} l^{\tilde t}_{11}
           Q^L_{11}{}' Q^R_{11}{}' m_b m_{{\tilde \chi}^0_1}
      + 2 \, Q^L_{11}{}' Q^R_{11}{}' \left( (k^{\tilde t}_{11})^2
              + (l^{\tilde t}_{11})^2 \right)
              m_{{\tilde \chi}^0_1} m_{\tilde{\chi}^+_1} \nonumber \\
  & & + \left( (k^{\tilde t}_{11})^2 (Q^R_{11}{}')^2
             + (l^{\tilde t}_{11})^2 (Q^L_{11}{}')^2 \right)
         \left(2 \, m^2_b+2 \, m^2_{{\tilde \chi}^0_1}
             +m^2_{H^+}+m^2_{{\tilde t}_1}\right)  \nonumber \\
  & & + 2 \, k^{\tilde t}_{11} l^{\tilde t}_{11}
          \left( (Q^L_{11}{}')^2 + (Q^R_{11}{}')^2 \right)
               m_b m_{\tilde{\chi}^+_1} \, , \\
a_{14} &=& - (k^{\tilde t}_{11})^2 (Q^R_{11}{}')^2 -
                 (l^{\tilde t}_{11})^2 (Q^L_{11}{}')^2  \, ,
\end{eqnarray}
\begin{eqnarray}
a_{31} &=& - 2  \left(l^{\tilde t}_{11} l^{\tilde t}_{12} Q^L_{12}{}'
            Q^R_{11}{}' + k^{\tilde t}_{11} k^{\tilde t}_{12} 
        Q^L_{11}{}' Q^R_{12}{}'\right) m_{{\tilde \chi}^0_1}
             m_{\tilde{\chi}^+_1}
         \left( m^2_{{\tilde \chi}^0_1} + m^2_{H^+} + 2 \, m^2_b \right)
        \nonumber \\
  & &   - 4 \, \left(k^{\tilde t}_{11} l^{\tilde t}_{12}
        Q^L_{12}{}' Q^R_{11}{}' +  k^{\tilde t}_{12} l^{\tilde t}_{11}
           Q^L_{11}{}' Q^R_{12}{}'\right) m_b m_{{\tilde \chi}^0_1}
      \left( m^2_{{\tilde \chi}^0_1} + m^2_{H^+} + m^2_b
            + m^2_{{\tilde t}_1} \right) \nonumber \\
  & &  - 4 \, \left(k^{\tilde t}_{12} l^{\tilde t}_{11}
           Q^L_{12}{}' Q^R_{11}{}' +  k^{\tilde t}_{11} l^{\tilde t}_{12}
           Q^L_{11}{}' Q^R_{12}{}'\right) m_b m_{{\tilde \chi}^0_1}
         m_{\tilde{\chi}^+_1} m_{\tilde{\chi}^+_2} \nonumber \\
  & &  - 2 \, \left(k^{\tilde t}_{11} k^{\tilde t}_{12}
     Q^L_{12}{}' Q^R_{11}{}' +   l^{\tilde t}_{11} l^{\tilde t}_{12}
     Q^L_{11}{}' Q^R_{12}{}'\right) m_{{\tilde \chi}^0_1}
      m_{\tilde{\chi}^+_2}
      \left( m^2_{{\tilde \chi}^0_1} + m^2_{H^+} + 2 \, m^2_b \right) 
      \nonumber \\
  & &  - 2 \, \left(l^{\tilde t}_{11} l^{\tilde t}_{12}
        Q^L_{11}{}' Q^L_{12}{}' +  k^{\tilde t}_{11} k^{\tilde t}_{12} 
        Q^R_{11}{}' Q^R_{12}{}'\right)     \nonumber \\
  & & \hspace{5mm} * \left[ \left(m^2_b+m^2_{{\tilde \chi}^0_1}\right)^2
      + \left(m^2_{{\tilde \chi}^0_1}+m^2_{{\tilde t}_1}\right)
        \left(m^2_b+m^2_{H^+}\right) \right] \nonumber \\
  & &  - 2 \, \left(k^{\tilde t}_{11} l^{\tilde t}_{12}
       Q^L_{11}{}' Q^L_{12}{}' +  k^{\tilde t}_{12} l^{\tilde t}_{11}
       Q^R_{11}{}' Q^R_{12}{}'\right) m_b m_{\tilde{\chi}^+_1}
     \left(m^2_b+m^2_{{\tilde t}_1}+2 \, m^2_{{\tilde \chi}^0_1} \right)
     \nonumber \\
  & &  - 2 \, \left(k^{\tilde t}_{12} l^{\tilde t}_{11}
     Q^L_{11}{}' Q^L_{12}{}' +  k^{\tilde t}_{11} l^{\tilde t}_{12}
     Q^R_{11}{}' Q^R_{12}{}'\right) m_b m_{\tilde{\chi}^+_2}
     \left(m^2_b+m^2_{{\tilde t}_1}+2 \, m^2_{{\tilde \chi}^0_1} \right)
     \nonumber \\
  & &  - 2 \, \left(k^{\tilde t}_{11} k^{\tilde t}_{12}
    Q^L_{11}{}' Q^L_{12}{}' +  l^{\tilde t}_{11} l^{\tilde t}_{12} Q^R_{11}{}'
    Q^R_{12}{}'\right) m_{\tilde{\chi}^+_1} m_{\tilde{\chi}^+_2}
    \left(m^2_{{\tilde \chi}^0_1} + m^2_b \right)  \, , \\
a_{32} &=& 2 \left(l^{\tilde t}_{11} l^{\tilde t}_{12} Q^L_{12}{}'
        Q^R_{11}{}' +   k^{\tilde t}_{11} k^{\tilde t}_{12} Q^L_{11}{}'
        Q^R_{12}{}'\right) m_{{\tilde \chi}^0_1} m_{\tilde{\chi}^+_1}
      \nonumber \\
  & &  + 4 \, \left(k^{\tilde t}_{11} l^{\tilde t}_{12}
      Q^L_{12}{}' Q^R_{11}{}' +  k^{\tilde t}_{12} l^{\tilde t}_{11}
      Q^L_{11}{}' Q^R_{12}{}'\right) m_b m_{{\tilde \chi}^0_1} \nonumber \\
  & &  + 2 \, \left(k^{\tilde t}_{11} k^{\tilde t}_{12} 
       Q^L_{12}{}' Q^R_{11}{}' +  l^{\tilde t}_{11} l^{\tilde t}_{12}
       Q^L_{11}{}' Q^R_{12}{}'\right) m_{{\tilde \chi}^0_1}
       m_{\tilde{\chi}^+_2} \nonumber \\
  & &  + 2 \, \left(l^{\tilde t}_{11} l^{\tilde t}_{12}
      Q^L_{11}{}' Q^L_{12}{}' +   k^{\tilde t}_{11} k^{\tilde t}_{12} 
      Q^R_{11}{}' Q^R_{12}{}'\right)
    \left( m^2_b+m^2_{{\tilde \chi}^0_1} \right)  \nonumber \\
  & &  + 2 \, \left(k^{\tilde t}_{11} l^{\tilde t}_{12}
      Q^L_{11}{}' Q^L_{12}{}' +  k^{\tilde t}_{12} l^{\tilde t}_{11}
      Q^R_{11}{}' Q^R_{12}{}'\right) m_b m_{\tilde{\chi}^+_1} 
      \nonumber \\
  & &  + 2 \, \left(k^{\tilde t}_{12} l^{\tilde t}_{11}
        Q^L_{11}{}' Q^L_{12}{}' +    k^{\tilde t}_{11} l^{\tilde t}_{12}
        Q^R_{11}{}' Q^R_{12}{}'\right) m_b m_{\tilde{\chi}^+_2}
      \nonumber \\
  & &  + 2 \, \left(k^{\tilde t}_{11} k^{\tilde t}_{12}
         Q^L_{11}{}' Q^L_{12}{}' +   l^{\tilde t}_{11} l^{\tilde t}_{12} 
         Q^R_{11}{}' Q^R_{12}{}'\right) m_{\tilde{\chi}^+_1}
          m_{\tilde{\chi}^+_2}  \, ,\\
a_{33} &=& 2 \left(l^{\tilde t}_{11} l^{\tilde t}_{12} Q^L_{12}{}'
          Q^R_{11}{}' +  k^{\tilde t}_{11} k^{\tilde t}_{12} Q^L_{11}{}'
          Q^R_{12}{}'\right) m_{{\tilde \chi}^0_1}
          m_{\tilde{\chi}^+_1} \nonumber \\
  & &  + 4 \, \left(k^{\tilde t}_{11} l^{\tilde t}_{12}
        Q^L_{12}{}' Q^R_{11}{}' +  k^{\tilde t}_{12} l^{\tilde t}_{11}
        Q^L_{11}{}' Q^R_{12}{}'\right) m_b m_{{\tilde \chi}^0_1} \nonumber \\
  & &  + 2 \, \left(k^{\tilde t}_{11} k^{\tilde t}_{12}
        Q^L_{12}{}' Q^R_{11}{}' +   l^{\tilde t}_{11} l^{\tilde t}_{12}
        Q^L_{11}{}' Q^R_{12}{}'\right) m_{{\tilde \chi}^0_1}
        m_{\tilde{\chi}^+_2} \nonumber \\
  & &  + 2 \, \left(l^{\tilde t}_{11} l^{\tilde t}_{12}
        Q^L_{11}{}' Q^L_{12}{}' +  k^{\tilde t}_{11} k^{\tilde t}_{12}
        Q^R_{11}{}' Q^R_{12}{}'\right)
   \left(2 \, m^2_b+2 \, m^2_{{\tilde \chi}^0_1}
         +m^2_{H^+}+m^2_{{\tilde t}_1} \right) \nonumber \\
  & &  + 2 \, \left(k^{\tilde t}_{11} l^{\tilde t}_{12}
      Q^L_{11}{}' Q^L_{12}{}' +  k^{\tilde t}_{12} l^{\tilde t}_{11}
     Q^R_{11}{}' Q^R_{12}{}'\right) m_b m_{\tilde{\chi}^+_1} \nonumber \\
  & &  + 2 \, \left(k^{\tilde t}_{12} l^{\tilde t}_{11}
      Q^L_{11}{}' Q^L_{12}{}' +  k^{\tilde t}_{11} l^{\tilde t}_{12}
      Q^R_{11}{}' Q^R_{12}{}'\right) m_b m_{\tilde{\chi}^+_2} \, , \\
a_{34} &=& - 2 \, \left(l^{\tilde t}_{11} l^{\tilde t}_{12} Q^L_{11}{}'
            Q^L_{12}{}' +  k^{\tilde t}_{11} k^{\tilde t}_{12} Q^R_{11}{}'
            Q^R_{12}{}'\right) \, ,
\end{eqnarray}
\begin{eqnarray}
b_{11} &=& \frac{\sqrt2 \,}{m_W} \Bigg\{
       b^{\tilde{t}}_{11} k^{\tilde t}_{11} Q^L_{11}{}'
       m_{\tilde{\chi}^+_1} m_b m_t
       \left[ \left( m^2_{{\tilde t}_1}-m^2_{{\tilde \chi}^0_1} \right)
       \cot \beta  - \left( m^2_b + m^2_{{\tilde \chi}^0_1} \right)
       \tan \beta \right] \nonumber \\
  & &  + b^{\tilde{t}}_{11} l^{\tilde t}_{11} Q^L_{11}{}' m_t
       \left[ \left( m^2_{H^+} m^2_{{\tilde t}_1} - m^2_b
              m^2_{{\tilde \chi}^0_1} \right) \cot \beta
            - m^2_b  \left( m^2_b + m^2_{{\tilde t}_1} + 2 \,
             m^2_{{\tilde \chi}^0_1} \right)
             \tan \beta \right] \nonumber \\
  & &  + a^{\tilde{t}}_{11} l^{\tilde t}_{11} Q^L_{11}{}'
           m_{{\tilde \chi}^0_1} \nonumber \\
  & & \hspace{2mm} * \left[ m^2_b
          \left( m^2_{H^+}+m^2_{{\tilde t}_1}-m^2_{{\tilde \chi}^0_1}
                  -m^2_b \right) \tan \beta
          - m^2_t \left( 2 \, m^2_b +m^2_{{\tilde \chi}^0_1}
                          +m^2_{H^+} \right)
          \cot \beta \right] \nonumber \\
  & &  + a^{\tilde{t}}_{11} k^{\tilde t}_{11} Q^L_{11}{}'
            m_{\tilde{\chi}^+_1} m_b m_{{\tilde \chi}^0_1}
       \left[ \left(m^2_{H^+}-m^2_b \right) \tan \beta
          - 2 \, m^2_t \cot \beta \right] \nonumber \\
  & &  + b^{\tilde{t}}_{11} k^{\tilde t}_{11} Q^R_{11}{}'
                m_{{\tilde \chi}^0_1} m_b m_t \nonumber \\
  & &  \hspace{2mm}* \left[ \left(m^2_{{\tilde t}_1}+m^2_{H^+}-m^2_b
                -m^2_{{\tilde \chi}^0_1} \right) \cot \beta
         - \left( m^2_{H^+}+2 \, m^2_b +m^2_{{\tilde \chi}^0_1} \right)
             \tan \beta \right] \nonumber \\
  & &  + b^{\tilde{t}}_{11} l^{\tilde t}_{11} Q^R_{11}{}'
                m_t m_{{\tilde \chi}^0_1} m_{\tilde{\chi}^+_1}
      \left[ \left(m^2_{H^+}-m^2_b \right) \cot \beta
         - 2 \, m^2_b  \tan \beta \right] \nonumber \\
  & &  - a^{\tilde{t}}_{11} k^{\tilde t}_{11} Q^R_{11}{}'
            m_b \nonumber \\
  & & \hspace{2mm} *   \left[ \left(m^2_b m^2_{{\tilde \chi}^0_1}
           - m^2_{{\tilde t}_1} m^2_{H^+} \right) \tan \beta
         + m^2_t \left( m^2_b +m^2_{{\tilde t}_1}+2 \,
             m^2_{{\tilde \chi}^0_1} \right)
          \cot \beta \right]  \nonumber \\
  & &  - a^{\tilde{t}}_{11} l^{\tilde t}_{11} Q^R_{11}{}'
              m_{\tilde{\chi}^+_1}
      \left[ m^2_b  \left(m^2_{{\tilde \chi}^0_1}-m^2_{{\tilde t}_1} \right)
           \tan \beta
         + m^2_t \left(m^2_b +m^2_{{\tilde \chi}^0_1} \right) \cot \beta
           \right] \Bigg\}   \, , \\
b_{12} &=& \frac{\sqrt2 \,}{m_W} \Bigg\{ b^{\tilde{t}}_{11}
                k^{\tilde t}_{11} Q^L_{11}{}'
              m_{\tilde{\chi}^+_1} m_b m_t \tan \beta
      + b^{\tilde{t}}_{11} l^{\tilde t}_{11} Q^L_{11}{}' m^2_b  m_t \tan \beta
              \nonumber \\
  & &  + a^{\tilde{t}}_{11} l^{\tilde t}_{11} Q^L_{11}{}'
              m^2_t m_{{\tilde \chi}^0_1} \cot \beta
              + b^{\tilde{t}}_{11} k^{\tilde t}_{11} Q^R_{11}{}'
            m_{{\tilde \chi}^0_1} m_b m_t \tan \beta \nonumber \\
  & &  + a^{\tilde{t}}_{11} l^{\tilde t}_{11} Q^R_{11}{}'
             m_{\tilde{\chi}^+_1} m^2_t \cot \beta
               + a^{\tilde{t}}_{11} k^{\tilde t}_{11} Q^R_{11}{}' m_b m^2_t
                 \cot \beta \Bigg\}  \, ,\\
b_{13} &=& - \frac{\sqrt2 \,}{m_W} \Bigg\{ b^{\tilde{t}}_{11}
             k^{\tilde t}_{11} Q^L_{11}{}'
           m_{\tilde{\chi}^+_1} m_b m_t \cot \beta
    - a^{\tilde{t}}_{11} l^{\tilde t}_{11} Q^L_{11}{}'
           m^2_t m_{{\tilde \chi}^0_1} \cot \beta \nonumber \\
  & &   + b^{\tilde{t}}_{11} l^{\tilde t}_{11} Q^L_{11}{}' m_t
     \left[ \left( m^2_{{\tilde \chi}^0_1}
            +m^2_{H^+}+m^2_{{\tilde t}_1}+m^2_b \right) \cot \beta
          - m^2_b  \tan \beta \right] \nonumber \\
  & &   + a^{\tilde{t}}_{11} k^{\tilde t}_{11}
  Q^L_{11}{}' m_{\tilde{\chi}^+_1} m_b m_{{\tilde \chi}^0_1} \tan \beta
    - b^{\tilde{t}}_{11} k^{\tilde t}_{11} Q^R_{11}{}' m_{{\tilde \chi}^0_1}
              m_b m_t \tan \beta \nonumber \\
  & &   + b^{\tilde{t}}_{11} l^{\tilde t}_{11} Q^R_{11}{}'
         m_{\tilde{\chi}^+_1} m_t m_{{\tilde \chi}^0_1} \cot \beta
   + a^{\tilde{t}}_{11} l^{\tilde t}_{11} Q^R_{11}{}' m^2_b 
            m_{\tilde{\chi}^+_1} \tan \beta \nonumber \\
  & &   + a^{\tilde{t}}_{11} k^{\tilde t}_{11} Q^R_{11}{}' m_b
     \left[ \left(m^2_b +m^2_{{\tilde \chi}^0_1} +m^2_{H^+}
                 +m^2_{{\tilde t}_1} \right) \tan \beta
          - m^2_t \cot \beta \right] \Bigg\}  \, ,\\
b_{14} &=& \frac{\sqrt2 \,}{m_W}
         \left( b^{\tilde{t}}_{11} l^{\tilde t}_{11}
                Q^L_{11}{}' m_t \cot \beta
               + a^{\tilde{t}}_{11} k^{\tilde t}_{11} Q^R_{11}{}'
                 m_b \tan \beta \right) \, ,
\end{eqnarray}
\begin{eqnarray}
c_{111} &=& 2 \, C^H_{{\tilde t}_1 {\tilde b}_1} \Bigg[
         \left(b^{\tilde b}_{11} l^{\tilde t}_{11} Q^L_{11}{}'
                        + a^{\tilde b}_{11} k^{\tilde t}_{11} Q^R_{11}{}'
         \right) m_b \left( m^2_b + m^2_{{\tilde t}_1} + 2 \,
             m^2_{{\tilde \chi}^0_1} \right) \nonumber \\
  & & \hspace{13mm} + \left(a^{\tilde b}_{11} l^{\tilde t}_{11} Q^L_{11}{}'
                        + b^{\tilde b}_{11} k^{\tilde t}_{11} Q^R_{11}{}'
                \right) m_{{\tilde \chi}^0_1}
           \left( m^2_{{\tilde \chi}^0_1} + m^2_{H^+} + 2 \, m^2_b
             \right) \nonumber \\
  & & \hspace{13mm} + \left(b^{\tilde b}_{11} k^{\tilde t}_{11} Q^L_{11}{}'
                        + a^{\tilde b}_{11} l^{\tilde t}_{11} Q^R_{11}{}'
     \right) m_{\tilde{\chi}^+_1} \left( m^2_b+m^2_{{\tilde \chi}^0_1}
        \right) \nonumber \\
  & & \hspace{13mm} + \left(a^{\tilde b}_{11} k^{\tilde t}_{11} Q^L_{11}{}'
                        + b^{\tilde b}_{11} l^{\tilde t}_{11} Q^R_{11}{}'
         \right) 2 \, m_b m_{{\tilde \chi}^0_1} m_{\tilde{\chi}^+_1}
      \Bigg]  \, ,\\
c_{112} &=& - 2 \, C^H_{{\tilde t}_1 {\tilde b}_1} \Bigg[
     \left(b^{\tilde b}_{11} l^{\tilde t}_{11} Q^L_{11}{}'
    + a^{\tilde b}_{11} k^{\tilde t}_{11} Q^R_{11}{}' \right) m_b
  + \left(a^{\tilde b}_{11} l^{\tilde t}_{11} Q^L_{11}{}'
      + b^{\tilde b}_{11} k^{\tilde t}_{11} Q^R_{11}{}' \right)
                m_{{\tilde \chi}^0_1} \nonumber \\
   & & \hspace{16mm} + \left(b^{\tilde b}_{11} k^{\tilde t}_{11} Q^L_{11}{}'
                        + a^{\tilde b}_{11} l^{\tilde t}_{11} Q^R_{11}{}' 
            \right) m_{\tilde{\chi}^+_1} \Bigg]  \, ,\\
c_{113} &=& -2 \, C^H_{{\tilde t}_1 {\tilde b}_1} \Bigg[
              \left(b^{\tilde b}_{11} l^{\tilde t}_{11} Q^L_{11}{}'
      + a^{\tilde b}_{11} k^{\tilde t}_{11} Q^R_{11}{}' \right) m_b
 + \left(a^{\tilde b}_{11} l^{\tilde t}_{11} Q^L_{11}{}'
      + b^{\tilde b}_{11} k^{\tilde t}_{11} Q^R_{11}{}' \right)
        m_{{\tilde \chi}^0_1} \Bigg] \, , \nonumber \\
\end{eqnarray}
\begin{eqnarray}
d_1 &=& \frac1{2 \, m^2_W} \Bigg[
   - \left( (a^{\tilde{t}}_{11})^2 m^2_t \cot^2 \beta
             + (b^{\tilde{t}}_{11})^2 m^2_b \tan^2 \beta \right)
       m^2_t \left( m^2_b + m^2_{{\tilde \chi}^0_1} \right) \nonumber \\
  & & \hspace{5mm} + \left( (a^{\tilde{t}}_{11})^2 m^2_b \tan^2 \beta
                     + (b^{\tilde{t}}_{11})^2 m^2_t \cot^2 \beta \right)
          \left( m^2_{{\tilde \chi}^0_1} - m^2_{{\tilde t}_1} \right)
          \left( m^2_{H^+} - m^2_b \right)  \nonumber \\
  & & \hspace{5mm} + 2 \, a^{\tilde{t}}_{11} b^{\tilde{t}}_{11}
             m_t m_{{\tilde \chi}^0_1}
      \left[ \left(m^2_{H^+}-m^2_b \right)
             \left(m^2_b \tan^2 \beta + m^2_t \cot^2 \beta \right)
        - 2 \, m^2_b m^2_t \right]  \nonumber \\
  & & \hspace{5mm} + 2 \, \left( (a^{\tilde{t}}_{11})^2 
             + (b^{\tilde{t}}_{11})^2 \right)
       m^2_b m^2_t \left( m^2_{{\tilde t}_1}
            - m^2_{{\tilde \chi}^0_1} \right) \Bigg]  \, , \\
d_2 &=& \frac{m^2_t}{2 \, m^2_W}
 \left( (a^{\tilde{t}}_{11})^2 m^2_t \cot^2 \beta
           + (b^{\tilde{t}}_{11})^2 m^2_b \tan^2 \beta \right)  \, , \\
d_{3} &=& \frac{-1}{2  m^2_W} \Bigg[
       2  \left( (a^{\tilde{t}}_{11})^2 + (b^{\tilde{t}}_{11})^2 \right)
         m^2_b m^2_t 
 + 2  a^{\tilde{t}}_{11} b^{\tilde{t}}_{11}
              m_t m_{{\tilde \chi}^0_1}
   \left( m^2_b \left( 2 +\tan^2 \beta \right)
            + m^2_t \cot^2 \beta \right) \nonumber \\
  & & \hspace{12mm}
 - \left( (a^{\tilde{t}}_{11})^2 m^2_b \tan^2 \beta
                     + (b^{\tilde{t}}_{11})^2 m^2_t \cot^2 \beta \right)
        \left(m^2_{H^+}+m^2_{{\tilde t}_1} \right) \Bigg]  \, ,\\
d_{4} &=& - \frac{(a^{\tilde{t}}_{11})^2 m^2_b \tan^2 \beta
                    + (b^{\tilde{t}}_{11})^2 m^2_t \cot^2 \beta}
                {2 \, m^2_W} \, ,
\end{eqnarray}
\begin{eqnarray}
e_{11} &=& - \frac{ \sqrt2 \, C^H_{{\tilde t}_1 {\tilde b}_1}}{m_W}
                \Bigg\{
       a^{\tilde{b}}_{11} b^{\tilde{t}}_{11} m_t m_{{\tilde \chi}^0_1}
       \left[ \left( m^2_{H^+} - m^2_b \right) \cot \beta
             - 2 \, m^2_b \tan \beta \right] \nonumber \\
  & & \hspace{21mm} + b^{\tilde{b}}_{11} b^{\tilde{t}}_{11} m_b m_t
      \left[ \left( m^2_{{\tilde t}_1} - m^2_{{\tilde \chi}^0_1} \right)
             \cot \beta
           - \left(m^2_b+m^2_{{\tilde \chi}^0_1} \right) \tan \beta \right]
         \nonumber \\
  & & \hspace{21mm} - a^{\tilde{b}}_{11} a^{\tilde{t}}_{11}
       \left[ m^2_t \left( m^2_b+m^2_{{\tilde \chi}^0_1} \right) \cot \beta
         + m^2_b \left(m^2_{{\tilde \chi}^0_1} -m^2_{{\tilde t}_1} \right)
              \tan \beta \right] \nonumber \\
  & & \hspace{21mm} - b^{\tilde{b}}_{11} a^{\tilde{t}}_{11}
              m_b m_{{\tilde \chi}^0_1}
       \left[ 2 \, m^2_t \cot \beta
             + \left( m^2_b-m^2_{H^+} \right) \tan \beta \right] \Bigg\}
 \, , \\
e_{12} &=& - \frac{ \sqrt2 \, C^H_{{\tilde t}_1 {\tilde b}_1}}{m_W}
       \left(
           a^{\tilde{b}}_{11} a^{\tilde{t}}_{11} m^2_t \cot \beta
      + b^{\tilde{b}}_{11} b^{\tilde{t}}_{11} m_b m_t \tan \beta \right)
 \, , \\
e_{13} &=& \frac{ \sqrt2 \, C^H_{{\tilde t}_1 {\tilde b}_1}}{m_W}
    \bigg(  a^{\tilde{b}}_{11} b^{\tilde{t}}_{11} m_t
       m_{{\tilde \chi}^0_1} \cot \beta
       + b^{\tilde{b}}_{11} b^{\tilde{t}}_{11} m_b m_t \cot \beta \nonumber \\
  & & \hspace{17mm} + \, a^{\tilde{b}}_{11} a^{\tilde{t}}_{11} m^2_b \tan \beta
       + b^{\tilde{b}}_{11} a^{\tilde{t}}_{11} m_b
              m_{{\tilde \chi}^0_1} \tan \beta \bigg)  \, ,
\end{eqnarray}
\begin{eqnarray}
f_{11} &=& - (C^H_{{\tilde t}_1 {\tilde b}_1})^2
         \left[ \left((a^{\tilde{b}}_{11})^2+(b^{\tilde{b}}_{11})^2 \right)
                 \left( m^2_b + m^2_{{\tilde \chi}^0_1} \right)
     + 4 \, a^{\tilde{b}}_{11} b^{\tilde{b}}_{11} m_b m_{{\tilde \chi}^0_1}
       \right]  \, ,\\
f_{12} &=& (C^H_{{\tilde t}_1 {\tilde b}_1})^2
        \left((a^{\tilde{b}}_{11})^2+(b^{\tilde{b}}_{11})^2 \right) \, ,
\end{eqnarray}
\begin{eqnarray}
f_{31} &=& - 2 \, C^H_{{\tilde t}_1 {\tilde b}_1}
                  C^H_{{\tilde t}_1 {\tilde b}_2}
 \left[ \left( a^{\tilde{b}}_{11} a^{\tilde{b}}_{12}
     + b^{\tilde{b}}_{11} b^{\tilde{b}}_{12} \right)
               \left( m^2_b + m^2_{{\tilde \chi}^0_1} \right)
     + 2  \left( a^{\tilde{b}}_{11} b^{\tilde{b}}_{12}
         + b^{\tilde{b}}_{11} a^{\tilde{b}}_{12} \right)
         m_b m_{{\tilde \chi}^0_1} \right]  \, ,\nonumber \\ \\
f_{32} &=& 2 \, C^H_{{\tilde t}_1 {\tilde b}_1}
        C^H_{{\tilde t}_1 {\tilde b}_2}
       \left( a^{\tilde{b}}_{11} a^{\tilde{b}}_{12}
              + b^{\tilde{b}}_{11} b^{\tilde{b}}_{12} \right) \, .
\end{eqnarray}
One gets the remaining coefficients by replacements:\\
\begin{tabular}{ll}
  $a_{1i} \to a_{2i}$: & $l^{\tilde t}_{11} \to l^{\tilde t}_{12}$,
            $k^{\tilde t}_{11} \to k^{\tilde t}_{12}$,
             $Q^L_{11}{}' \to Q^L_{12}{}'$, $Q^R_{11}{}' \to Q^R_{12}{}'$,
                          $m_{{\tilde \chi}^+_1} \to m_{{\tilde \chi}^+_2}$ \\
  $b_{1i} \to b_{2i}$: & $l^{\tilde t}_{11} \to l^{\tilde t}_{12}$,
   $k^{\tilde t}_{11} \to k^{\tilde t}_{12}$,
       $Q^L_{11}{}' \to Q^L_{12}{}'$, $Q^R_{11}{}' \to Q^R_{12}{}'$,
                      $m_{{\tilde \chi}^+_1} \to m_{{\tilde \chi}^+_2}$ \\
  $c_{11i} \to c_{12i}$: & $l^{\tilde t}_{11} \to l^{\tilde t}_{12}$,
           $k^{\tilde t}_{11} \to k^{\tilde t}_{12}$,
            $Q^L_{11}{}' \to Q^L_{12}{}'$, $Q^R_{11}{}' \to Q^R_{12}{}'$,
             $m_{{\tilde \chi}^+_1} \to m_{{\tilde \chi}^+_2}$ \\
  $c_{11i} \to c_{21i}$: & $a^{\tilde{b}}_{11} \to a^{\tilde{b}}_{12}$,
                            $b^{\tilde{b}}_{11} \to b^{\tilde{b}}_{12}$,
                            $m_{{\tilde b}_1} \to m_{{\tilde b}_2}$,
                            $C^H_{{\tilde t}_1 {\tilde b}_1} \to
                             C^H_{{\tilde t}_1 {\tilde b}_2}$ \\
  $c_{11i} \to c_{22i}$: & $l^{\tilde t}_{11} \to l^{\tilde t}_{12}$,
           $k^{\tilde t}_{11} \to k^{\tilde t}_{12}$,
          $Q^L_{11}{}' \to Q^L_{12}{}'$, $Q^R_{11}{}' \to Q^R_{12}{}'$,
                       $m_{{\tilde \chi}^+_1} \to m_{{\tilde \chi}^+_2}$, \\
                          & $a^{\tilde{b}}_{11} \to a^{\tilde{b}}_{12}$,
                            $b^{\tilde{b}}_{11} \to b^{\tilde{b}}_{12}$,
                            $m_{{\tilde b}_1} \to m_{{\tilde b}_2}$,
                          $C^H_{{\tilde t}_1 {\tilde b}_1} \to
                           C^H_{{\tilde t}_1 {\tilde b}_2}$ \\
  $e_{1i} \to e_{2i}$: & $a^{\tilde{b}}_{11} \to a^{\tilde{b}}_{12}$,
                          $b^{\tilde{b}}_{11} \to b^{\tilde{b}}_{12}$,
                          $m_{{\tilde b}_1} \to m_{{\tilde b}_2}$,
                          $C^H_{{\tilde t}_1 {\tilde b}_1} \to
                           C^H_{{\tilde t}_1 {\tilde b}_2}$ \\
  $f_{1i} \to f_{2i}$: & $a^{\tilde{b}}_{11} \to a^{\tilde{b}}_{12}$,
                          $b^{\tilde{b}}_{11} \to b^{\tilde{b}}_{12}$,
                          $m_{{\tilde b}_1} \to m_{{\tilde b}_2}$,
                          $C^H_{{\tilde t}_1 {\tilde b}_1} \to
                           C^H_{{\tilde t}_1 {\tilde b}_2}$ \\
\end{tabular}

\subsection{The widths $\Gamma({\tilde t}_1 \to b \, \tilde{l} \, l')$}

Here the decay width is given by
\begin{eqnarray}
 \Gamma({\tilde t}_1 \to b \, \tilde{l} \, l') =
 \frac{\alpha^2}  {16 \, \pi m^3_{{\tilde t}_1} \sin^4 \theta_W}
 \int\limits^{(m_{{\tilde t}_1}-m_b)^2}_{(m_{l'} + m_{\tilde{l}})^2} d \, s \,
        W_{l' \tilde{l}}(s)
        \sum^3_{i=1} \left( \sum^5_{j=1} c_{ij} s^{(j-4)} \right) D_i (s)
\end{eqnarray}
with
\begin{eqnarray}
 D_{1,2} (s) &=&
  \frac1{(s- m^2_{\tilde{\chi}^+_{1,2}})^2
           +  m^2_{\tilde{\chi}^+_{1,2}}
                \Gamma^2_{{\tilde \chi}^+_{1,2}}}  \, ,\\
 D_3 (s) &=& \mbox{Re} \left(
  \frac1{(s- m^2_{\tilde{\chi}^+_1} 
               + i m_{\tilde{\chi}^+_1} \Gamma_{{\tilde \chi}^+_1})
            (s- m^2_{\tilde{\chi}^+_2}
    - i m_{\tilde{\chi}^+_2} \Gamma_{{\tilde \chi}^+_2}) }  \right) \, .
\end{eqnarray}
In the case of ${\tilde t}_1 \to b \, \tilde{\nu}_e \, e^+$ one finds in the
limit $m_e \to 0$ that
\begin{eqnarray}
W_{e \tilde{\nu}_e}(s) &=& \lambda^{\frac{1}{2}}(s,m^2_{{\tilde t}_1},m^2_b)
                            \left(s-m^2_{\tilde{\nu}_e} \right) \, ,
\end{eqnarray}
\begin{eqnarray}
c_{11} &=& \frac{1}{2} (k^{\tilde t}_{11})^2 V^2_{11}
              m^2_{\tilde{\chi}^+_1} m^2_{\tilde{\nu}_e}
            \left( m^2_b-m^2_{{\tilde t}_1} \right) \, , \\
c_{12} &=& V^2_{11} \bigg[
      \frac{1}{2} (l^{\tilde t}_{11})^2 m^2_{\tilde{\nu}_e}
          \left( m^2_b - m^2_{{\tilde t}_1} \right)
    + \frac{1}{2} (k^{\tilde t}_{11})^2  m^2_{\tilde{\chi}^+_1}
             \left(m^2_{{\tilde t}_1}+m^2_{\tilde{\nu}_e}-m^2_b \right)
          \nonumber \\
    & & \hspace{8mm} + 2 \, k^{\tilde t}_{11} l^{\tilde t}_{11}
               m_b m_{\tilde{\chi}^+_1} m^2_{\tilde{\nu}_e}
       \bigg]  \, ,\\
c_{13} &=& V^2_{11} \left[ \frac{1}{2} (l^{\tilde t}_{11})^2
             \left( m^2_{{\tilde t}_1} + m^2_{\tilde{\nu}_e} - m^2_b \right)
        - 2 \, k^{\tilde t}_{11} l^{\tilde t}_{11}
                m_b m_{\tilde{\chi}^+_1}
  - \frac{1}{2} (k^{\tilde t}_{11})^2  m^2_{\tilde{\chi}^+_1} \right]
 \, ,  \\
c_{14} &=& - \frac{(l^{\tilde t}_{11})^2 V^2_{11}}2  \, ,
\end{eqnarray}
\begin{eqnarray}
c_{31} &=& k^{\tilde t}_{11} k^{\tilde t}_{12} V_{11}
            V_{12} m_{\tilde{\chi}^+_1} m_{\tilde{\chi}^+_2}
            (m^2_b-m^2_{{\tilde t}_1}) m^2_{\tilde{\nu}_e}  \, ,       \\
c_{32} &=& V_{11} V_{12} \bigg[
       l^{\tilde t}_{11} l^{\tilde t}_{12} m^2_{\tilde{\nu}_e}
              \left( m^2_b - m^2_{{\tilde t}_1} \right)
       + k^{\tilde t}_{11} k^{\tilde t}_{12}
                 m_{\tilde{\chi}^+_1} m_{\tilde{\chi}^+_2}
      \left( m^2_{{\tilde t}_1}+m^2_{\tilde{\nu}_e}-m^2_b \right) \nonumber \\
    & & \hspace{12mm} + 2 \, k^{\tilde t}_{11} l^{\tilde t}_{12}
     m_b m_{\tilde{\chi}^+_1} m^2_{\tilde{\nu}_e}
        + 2 \, k^{\tilde t}_{12} l^{\tilde t}_{11}
              m_b m_{\tilde{\chi}^+_2} m^2_{\tilde{\nu}_e} \bigg] \, , \\
c_{33} &=& V_{11} V_{12} \bigg[
        l^{\tilde t}_{11} l^{\tilde t}_{12}
            \left(m^2_{{\tilde t}_1}+m^2_{\tilde{\nu}_e}-m^2_b \right)
     - 2 \, k^{\tilde t}_{11} l^{\tilde t}_{12} m_b m_{\tilde{\chi}^+_1}
         \nonumber \\
    & & \hspace{12mm}  - 2 \, k^{\tilde t}_{12} l^{\tilde t}_{11}
               m_b m_{\tilde{\chi}^+_2}
         -  k^{\tilde t}_{11} k^{\tilde t}_{12} m_{\tilde{\chi}^+_1}
                  m_{\tilde{\chi}^+_2} \bigg]  \, , \\
c_{34} &=& - l^{\tilde t}_{11} l^{\tilde t}_{12} V_{11} V_{12}  \, , \\
c_{15} &=& c_{25} = c_{35} = 0 \, .
\end{eqnarray}
The coefficients $c_{2i}$ are obtained from $c_{1i}$ by the  
replacements:
$k^{\tilde t}_{11} \to k^{\tilde t}_{12},l^{\tilde t}_{11} \to
                l^{\tilde t}_{12},V_{11} \to V_{12}$ and
$m_{\tilde{\chi}^+_1} \to m_{\tilde{\chi}^+_2}$.
In the case of ${\tilde t}_1 \to b \, {\tilde \nu}_\tau  \, \tau^+$ one 
finds that
\begin{eqnarray}
W_{\tau {\tilde \nu}_\tau }(s) &=& \lambda^{\frac{1}{2}}
           (s,m^2_{{\tilde t}_1},m^2_b)
       \lambda^{\frac{1}{2}}(s,m^2_{{\tilde \nu}_\tau},m^2_\tau) \, ,
\end{eqnarray}
\begin{eqnarray}
c_{11} &=& \frac{1}{2} \left( (k^{{\tilde \nu}_\tau}_1)^2
       (l^{\tilde t}_{11})^2  + (l^{{\tilde \nu}_\tau}_1)^2
         (k^{\tilde t}_{11})^2 \right)  m^2_{\tilde{\chi}^+_1}
      \left( m^2_b-m^2_{{\tilde t}_1} \right) 
        \left(m_{{\tilde \nu}_\tau}^2-m^2_\tau \right)  \, ,\\
c_{12} &=& \frac{1}{2}
    \left( (k^{{\tilde \nu}_\tau}_1)^2 (k^{\tilde t}_{11})^2
          + (l^{{\tilde \nu}_\tau}_1)^2 (l^{\tilde t}_{11})^2 \right)
         \left(m^2_{{\tilde t}_1}-m^2_b \right)
       \left(m^2_\tau - m_{{\tilde \nu}_\tau}^2 \right) \nonumber \\
  & & + \frac{1}{2} \left( (k^{{\tilde \nu}_\tau}_1)^2
      (l^{\tilde t}_{11})^2   + (l^{{\tilde \nu}_\tau}_1)^2
           (k^{\tilde t}_{11})^2 \right)  m^2_{\tilde{\chi}^+_1}
        \left(m^2_{{\tilde t}_1}+m_{{\tilde \nu}_\tau}^2-m^2_b-m^2_\tau 
       \right) \nonumber \\
  & &  + 2 \, k^{{\tilde \nu}_\tau}_1 l^{{\tilde \nu}_\tau}_1
        \left((k^{\tilde t}_{11})^2+(l^{\tilde t}_{11})^2 \right)
          m_{\tau} m_{\tilde{\chi}^+_1}
              \left( m^2_{{\tilde t}_1}-m^2_b \right)  \nonumber \\
  & &  + 2 \, k^{\tilde t}_{11} l^{\tilde t}_{11}
          \left((k^{{\tilde \nu}_\tau}_1)^2+(l^{{\tilde \nu}_\tau}_1)^2
         \right) m_b m_{\tilde{\chi}^+_1} 
      \left(m_{{\tilde \nu}_\tau}^2-m^2_\tau \right) 
 - 4 \, k^{{\tilde \nu}_\tau}_1
        l^{{\tilde \nu}_\tau}_1 k^{\tilde t}_{11} l^{\tilde t}_{11}
          m_b m_{\tau}  m^2_{\tilde{\chi}^+_1}  \, , \\ 
c_{13} &=&  \frac{1}{2} \left( (k^{{\tilde \nu}_\tau}_1)^2
        (k^{\tilde t}_{11})^2  + (l^{{\tilde \nu}_\tau}_1)^2
             (l^{\tilde t}_{11})^2 \right)
    \left( m^2_{{\tilde t}_1} + m_{{\tilde \nu}_\tau}^2 - m^2_b - m^2_\tau 
        \right) \nonumber \\
    & &  - 2 \, k^{\tilde t}_{11} l^{\tilde t}_{11}
        \left((k^{{\tilde \nu}_\tau}_1)^2+(l^{{\tilde \nu}_\tau}_1)^2
               \right) m_b m_{\tilde{\chi}^+_1}
 - \frac{1}{2} \left( (k^{{\tilde \nu}_\tau}_1)^2
        (l^{\tilde t}_{11})^2  + (l^{{\tilde \nu}_\tau}_1)^2
   (k^{\tilde t}_{11})^2 \right)  m^2_{\tilde{\chi}^+_1} \nonumber \\
    & & \hspace{1mm} - 4 \, k^{{\tilde \nu}_\tau}_1 l^{{\tilde \nu}_\tau}_1
       k^{\tilde t}_{11} l^{\tilde t}_{11} m_b m_{\tau}
       - 2 \, k^{{\tilde \nu}_\tau}_1 l^{{\tilde \nu}_\tau}_1
        \left((k^{\tilde t}_{11})^2+(l^{\tilde t}_{11})^2 \right)
         m_\tau m_{\tilde{\chi}^+_1}  \, ,   \\
c_{14} &=& - \frac{1}{2} \left( (k^{{\tilde \nu}_\tau}_1)^2 
     (k^{\tilde t}_{11})^2  + (l^{{\tilde \nu}_\tau}_1)^2
         (l^{\tilde t}_{11})^2 \right) \, , 
\end{eqnarray}
\begin{eqnarray}
c_{31} &=& \left( k^{{\tilde \nu}_\tau}_1 k^{{\tilde \nu}_\tau}_2
 l^{\tilde t}_{11} l^{\tilde t}_{12}
                            + k^{\tilde t}_{11} k^{\tilde t}_{12} 
l^{{\tilde \nu}_\tau}_1 l^{{\tilde \nu}_\tau}_2 \right)
 m_{\tilde{\chi}^+_1} m_{\tilde{\chi}^+_2}
      \left( m^2_b-m^2_{{\tilde t}_1} \right)
 \left( m_{{\tilde \nu}_\tau}^2-m^2_\tau \right)  \, , \\
c_{32} &=&  \left( k^{{\tilde \nu}_\tau}_1 k^{{\tilde \nu}_\tau}_2
 k^{\tilde t}_{11} k^{\tilde t}_{12}   + l^{{\tilde \nu}_\tau}_1
 l^{{\tilde \nu}_\tau}_2 l^{\tilde t}_{11}  l^{\tilde t}_{12} \right)
 \left( m^2_{{\tilde t}_1}-m^2_b \right)
                   \left( m^2_\tau-m_{{\tilde \nu}_\tau}^2 \right) \nonumber \\
 & &  +  \left( k^{{\tilde \nu}_\tau}_1 k^{{\tilde \nu}_\tau}_2
     l^{\tilde t}_{11} l^{\tilde t}_{12}
                            + k^{\tilde t}_{11} k^{\tilde t}_{12}
 l^{{\tilde \nu}_\tau}_1 l^{{\tilde \nu}_\tau}_2 \right)
  m_{\tilde{\chi}^+_1} m_{\tilde{\chi}^+_2}
 \left( m^2_{{\tilde t}_1}+m_{{\tilde \nu}_\tau}^2-m^2_b-m^2_\tau \right)
    \nonumber \\
 & &  + 2  \left( k^{{\tilde \nu}_\tau}_1
        l^{{\tilde \nu}_\tau}_2 l^{\tilde t}_{11} l^{\tilde t}_{12}
     + k^{{\tilde \nu}_\tau}_2 k^{\tilde t}_{11} k^{\tilde t}_{12}
      l^{{\tilde \nu}_\tau}_1 \right) m_{\tau} m_{\tilde{\chi}^+_1}
               \left( m^2_{{\tilde t}_1}-m^2_b \right) \nonumber \\
 & &  + 2  \left( k^{{\tilde \nu}_\tau}_1 k^{\tilde t}_{11}
    k^{\tilde t}_{12} l^{{\tilde \nu}_\tau}_2
  + k^{{\tilde \nu}_\tau}_2 l^{{\tilde \nu}_\tau}_1 l^{\tilde t}_{11}
      l^{\tilde t}_{12} \right) m_{\tau} m_{\tilde{\chi}^+_2}
               \left( m^2_{{\tilde t}_1}-m^2_b \right) \nonumber \\
 & &  + 2  \left( k^{{\tilde \nu}_\tau}_1
          k^{{\tilde \nu}_\tau}_2 k^{\tilde t}_{12} l^{\tilde t}_{11}
      + k^{\tilde t}_{11} l^{{\tilde \nu}_\tau}_1 l^{{\tilde \nu}_\tau}_2
        l^{\tilde t}_{12} \right) m_b m_{\tilde{\chi}^+_1}
               \left( m_{{\tilde \nu}_\tau}^2-m^2_\tau \right) \nonumber \\
 & &  + 2  \left( k^{{\tilde \nu}_\tau}_1
         k^{{\tilde \nu}_\tau}_2 k^{\tilde t}_{11} l^{\tilde t}_{12}
       + k^{\tilde t}_{12} l^{{\tilde \nu}_\tau}_1 l^{{\tilde \nu}_\tau}_2
         l^{\tilde t}_{11} \right) m_b m_{\tilde{\chi}^+_2}
               \left( m_{{\tilde \nu}_\tau}^2-m^2_\tau \right) \nonumber \\
 & &  - 4  \left( k^{{\tilde \nu}_\tau}_1 k^{\tilde t}_{12}
         l^{{\tilde \nu}_\tau}_2 l^{\tilde t}_{11}
        + k^{{\tilde \nu}_\tau}_2 k^{\tilde t}_{11} l^{{\tilde \nu}_\tau}_1
          l^{\tilde t}_{12} \right) m_b m_{\tau} m_{\tilde{\chi}^+_1}
         m_{\tilde{\chi}^+_2}  \, , \\
c_{33} &=&  \left( k^{{\tilde \nu}_\tau}_1 k^{{\tilde \nu}_\tau}_2
          k^{\tilde t}_{11} k^{\tilde t}_{12}
      + l^{{\tilde \nu}_\tau}_1 l^{{\tilde \nu}_\tau}_2 l^{\tilde t}_{11}
       l^{\tilde t}_{12} \right)
     \left( m^2_{{\tilde t}_1}+m_{{\tilde \nu}_\tau}^2-m^2_b-m^2_\tau \right)
         \nonumber \\
 & & - 2 \left( k^{{\tilde \nu}_\tau}_1
       k^{{\tilde \nu}_\tau}_2
       k^{\tilde t}_{12} l^{\tilde t}_{11} + k^{\tilde t}_{11}
         l^{{\tilde \nu}_\tau}_1 l^{{\tilde \nu}_\tau}_2 l^{\tilde t}_{12}
       \right) m_b m_{\tilde{\chi}^+_1}
 - 2 \left( k^{{\tilde \nu}_\tau}_1
     k^{{\tilde \nu}_\tau}_2 k^{\tilde t}_{11} l^{\tilde t}_{12}
    + k^{\tilde t}_{12} l^{{\tilde \nu}_\tau}_1 l^{{\tilde \nu}_\tau}_2
    l^{\tilde t}_{11} \right) m_b m_{\tilde{\chi}^+_2} \nonumber \\
 & & - \left( k^{{\tilde \nu}_\tau}_1 k^{{\tilde \nu}_\tau}_2
      l^{\tilde t}_{11} l^{\tilde t}_{12}
     + k^{\tilde t}_{11} k^{\tilde t}_{12} l^{{\tilde \nu}_\tau}_1
        l^{{\tilde \nu}_\tau}_2 \right) m_{\tilde{\chi}^+_1}
            m_{\tilde{\chi}^+_2} 
  - 4  \left( k^{{\tilde \nu}_\tau}_1 k^{\tilde t}_{11}
        l^{{\tilde \nu}_\tau}_2 l^{\tilde t}_{12}
    + k^{{\tilde \nu}_\tau}_2 k^{\tilde t}_{12} l^{{\tilde \nu}_\tau}_1
           l^{\tilde t}_{11} \right) m_b m_{\tau} \nonumber \\
 & & 
 - 2  \left( k^{{\tilde \nu}_\tau}_1
          l^{{\tilde \nu}_\tau}_2 l^{\tilde t}_{11} l^{\tilde t}_{12}
        + k^{{\tilde \nu}_\tau}_2 k^{\tilde t}_{11} k^{\tilde t}_{12}
           l^{{\tilde \nu}_\tau}_1 \right) m_{\tau} m_{\tilde{\chi}^+_1}
 - 2 \left( k^{{\tilde \nu}_\tau}_1 k^{\tilde t}_{11}
        k^{\tilde t}_{12} l^{{\tilde \nu}_\tau}_2
       + k^{{\tilde \nu}_\tau}_2 l^{{\tilde \nu}_\tau}_1 l^{\tilde t}_{11}
      l^{\tilde t}_{12} \right) m_{\tau} m_{\tilde{\chi}^+_2} 
  \nonumber \\ \\
c_{34} &=& - \left( k^{{\tilde \nu}_\tau}_1 k^{{\tilde \nu}_\tau}_2
         k^{\tilde t}_{11} k^{\tilde t}_{12}
    + l^{{\tilde \nu}_\tau}_1 l^{{\tilde \nu}_\tau}_2 l^{\tilde t}_{11}
         l^{\tilde t}_{12} \right)  \, , \\
c_{15} &=& c_{25} = c_{35} = 0 \, .
\end{eqnarray}
The coefficients $c_{2i}$ are obtained from $c_{1i}$ by the  
replacements:
$k^{\tilde t}_{11} \to k^{\tilde t}_{12},l^{\tilde t}_{11}
      \to l^{\tilde t}_{12}, 
k^{{\tilde \nu}_\tau}_1 \to k^{{\tilde \nu}_\tau}_2,
l^{{\tilde \nu}_\tau}_1 \to l^{{\tilde \nu}_\tau}_2$
and $m_{\tilde{\chi}^+_1} \to m_{\tilde{\chi}^+_2}$.
In the case of ${\tilde t}_1 \to b \, {\tilde \tau}^+_1 \, \nu_\tau$ one finds:
\begin{eqnarray}
W_{\nu_\tau {\tilde \tau}_1}(s) &=&
 \lambda^{\frac{1}{2}}(s,m^2_{{\tilde t}_1},m^2_b) \, , 
\end{eqnarray}
\begin{eqnarray}
c_{11} &=& \frac{1}{2} (l^{\tilde \tau}_{11})^2 (l^{\tilde t}_{11})^2
    m^2_{\tilde{\chi}^+_1} m^4_{{\tilde \tau}_1}
            \left( m^2_{{\tilde t}_1}-m^2_b \right)  \, ,  \\
c_{12} &=& (l^{\tilde \tau}_{11})^2 \bigg[ (l^{\tilde t}_{11})^2
       m^2_{{\tilde \tau}_1}  m^2_{\tilde{\chi}^+_1}
                 \left( m^2_b-m^2_{{\tilde t}_1}-\frac{1}{2}
        m^2_{{\tilde \tau}_1} \right)
       + \frac{1}{2} (k^{\tilde t}_{11})^2 m^4_{{\tilde \tau}_1}
          \left( m^2_{{\tilde t}_1}-m^2_b \right)  \nonumber \\
  & & \hspace{10mm} - 2 \, k^{\tilde t}_{11} l^{\tilde t}_{11} m_b
         m_{\tilde{\chi}^+_1} m^4_{{\tilde \tau}_1} \bigg] \, ,  \\
c_{13} &=& (l^{\tilde \tau}_{11})^2 \bigg[ \frac{1}{2} (l^{\tilde t}_{11})^2 
 m^2_{\tilde{\chi}^+_1}
  \left( m^2_{{\tilde t}_1} + 2 \, m^2_{{\tilde \tau}_1} - m^2_b \right)
        + 4 \, k^{\tilde t}_{11} l^{\tilde t}_{11} m_b
           m_{\tilde{\chi}^+_1} m^2_{{\tilde \tau}_1}      \nonumber \\
  & & \hspace{10mm} + (k^{\tilde t}_{11})^2 m^2_{{\tilde \tau}_1}
            \left( m^2_b - m^2_{{\tilde t}_1} - \frac{1}{2}
        m^2_{{\tilde \tau}_1} \right) \bigg] \, ,  \\
c_{14} &=& (l^{\tilde \tau}_{11})^2 \bigg[ \frac{1}{2} (k^{\tilde t}_{11})^2
         \left( 2 \, m^2_{{\tilde \tau}_1} + m^2_{{\tilde t}_1} - m^2_b \right)
     - \frac{1}{2} (l^{\tilde t}_{11})^2  m^2_{\tilde{\chi}^+_1}
     - 2 \, k^{\tilde t}_{11} l^{\tilde t}_{11} m_b m_{\tilde{\chi}^+_1}
                 \bigg]   \, , \\ 
c_{15} &=& - \frac{1}{2} (l^{\tilde \tau}_{11})^2 (k^{\tilde t}_{11})^2 \, , 
\end{eqnarray}
\begin{eqnarray}
c_{31} &=& l^{\tilde \tau}_{11} l^{\tilde \tau}_{12} l^{\tilde t}_{11}
        l^{\tilde t}_{12} m_{\tilde{\chi}^+_1} m_{\tilde{\chi}^+_2}
      m^4_{{\tilde \tau}_1} \left( m^2_{{\tilde t}_1}-m^2_b \right) \, ,  \\
c_{32} &=& l^{\tilde \tau}_{11} l^{\tilde \tau}_{12} \bigg[ k^{\tilde t}_{11}
       k^{\tilde t}_{12} m^4_{{\tilde \tau}_1}
         \left( m^2_{{\tilde t}_1}-m^2_b \right)
     + 2 \, l^{\tilde t}_{11} l^{\tilde t}_{12} m_{\tilde{\chi}^+_1}
         m_{\tilde{\chi}^+_2} m^2_{{\tilde \tau}_1}
               \left( m^2_b - \frac{1}{2} m^2_{{\tilde \tau}_1}
            - m^2_{{\tilde t}_1} \right)  \nonumber \\
  & & \hspace{10mm} - 2 \, l^{\tilde t}_{11} k^{\tilde t}_{12} m_b
         m_{\tilde{\chi}^+_1} m^4_{{\tilde \tau}_1}
     - 2 \, l^{\tilde t}_{12} k^{\tilde t}_{11} m_b m_{\tilde{\chi}^+_2}
                m^4_{{\tilde \tau}_1}  \bigg]  \, , \\
c_{33} &=& l^{\tilde \tau}_{11} l^{\tilde \tau}_{12} \bigg[ l^{\tilde t}_{11}
              l^{\tilde t}_{12} m_{\tilde{\chi}^+_1}
        m_{\tilde{\chi}^+_2} \left( 2 \, m^2_{{\tilde \tau}_1} +
                  m^2_{{\tilde t}_1}-m^2_b \right)
     + 4 \, l^{\tilde t}_{11} k^{\tilde t}_{12} m_b m_{\tilde{\chi}^+_1}
                   m^2_{{\tilde \tau}_1}  \nonumber \\
   & & \hspace{10mm} + 4 \, l^{\tilde t}_{12} k^{\tilde t}_{11} m_b 
             m_{\tilde{\chi}^+_2} m^2_{{\tilde \tau}_1}
       - 2 \, k^{\tilde t}_{11} k^{\tilde t}_{12} m^2_{{\tilde \tau}_1}
             \left( m^2_{{\tilde t}_1} + \frac{1}{2} m^2_{{\tilde \tau}_1}
                  - m^2_b \right) \bigg]  \, , \\
c_{34} &=& l^{\tilde \tau}_{11} l^{\tilde \tau}_{12} \bigg[ k^{\tilde t}_{11}
                k^{\tilde t}_{12}
         \left( 2 \, m^2_{{\tilde \tau}_1} + m^2_{{\tilde t}_1}-m^2_b \right)
     - 2 \, l^{\tilde t}_{11} k^{\tilde t}_{12} m_b m_{\tilde{\chi}^+_1}
              \nonumber \\
   & & \hspace{10mm}  - 2 \, l^{\tilde t}_{12} k^{\tilde t}_{11} m_b
                  m_{\tilde{\chi}^+_2}
     - l^{\tilde t}_{11} l^{\tilde t}_{12} m_{\tilde{\chi}^+_1}
                m_{\tilde{\chi}^+_2}   \bigg]  \, ,   \\
c_{35} &=& - l^{\tilde \tau}_{11} l^{\tilde \tau}_{12} k^{\tilde t}_{11}
                    k^{\tilde t}_{12} \, .
\end{eqnarray}
The coefficients $c_{2i}$ are obtained from $c_{1i}$ by the  
replacements:
$k^{\tilde t}_{11} \to k^{\tilde t}_{12},l^{\tilde t}_{11} \to
       l^{\tilde t}_{12},l^{\tilde \tau}_{11} \to l^{\tilde \tau}_{12}$ and
$m_{\tilde{\chi}^+_1} \to m_{\tilde{\chi}^+_2}$.
To get the coefficients for ${\tilde t}_1 \to b \, {\tilde \tau}_2 \,
       \nu_\tau$ one has to
make the  replacements: $l^{\tilde \tau}_{1i} \to
 l^{\tilde \tau}_{2i}$ and 
$m_{{\tilde \tau}_1} \to m_{{\tilde \tau}_2}$.
For ${\tilde t}_1 \to b \, \tilde{e}^+_L \, \nu_e$ one gets the corresponding 
coefficients
by the replacements: $l^{\tilde \tau}_{1i} \to u_{1i} $ and
$m_{{\tilde \tau}_1} \to m_{\tilde{e}_L}$.

\section{Couplings}
\label{appB}

Here we give the couplings that were used in Eq.~(\ref{eq:interaction}):
The Yukawa couplings of the sfermions are given by:
\begin{eqnarray}
Y_{\tau} = \frac{m_{\tau}}{\sqrt2 \, m_W \cos \beta}, \hspace{5mm}
Y_{b} = \frac{m_{b}}{\sqrt2 \, m_W \cos \beta}, \hspace{5mm}
Y_{t} = \frac{m_{t}}{\sqrt2 \, m_W \sin \beta}.
\end{eqnarray}
The ${\tilde q}_i$-$q'$-${\tilde \chi}^\pm_j$ couplings read then
\begin{eqnarray}
l^{\tilde q}_{ij} = {\cal{R}}^{\tilde q}_{in} {\cal{O}}^{q}_{jn}, \hspace{5mm}
k^{\tilde q}_{ij} = {\cal{R}}^{\tilde q}_{i1} {\cal{O}}^{q'}_{j2}
\label{eq:kte}
\end{eqnarray}
with
\begin{eqnarray}
 {\cal{O}}^{t}_j = \left( \begin{array}{c} -V_{j1}
                      \\ Y_t V_{j2} \end{array} \right), \hspace{5mm}
 {\cal{O}}^{b}_j = \left( \begin{array}{c} -U_{j1}
                      \\ Y_b U_{j2} \end{array} \right) .
\end{eqnarray}
where $U_{ij}$ and $V_{ij}$ are the mixing matrices of the charginos
\cite{Bartl92a}. In case of sleptons we have:
\begin{eqnarray}
l^{\tilde \tau}_{ij} =
    {\cal{R}}^{\tilde \tau}_{in} {\cal{O}}^{\tau}_{jn}, \hspace{5mm}
k^{\tilde \tau}_{ij} = 0, \hspace{5mm}
l^{\tilde \nu}_j = - V_{j1}, \hspace{5mm}
k^{\tilde \nu}_j = Y_{\tau} U_{j2} \, ,
\end{eqnarray}
with
\begin{eqnarray}
 {\cal{O}}^{\tau}_j = \left( \begin{array}{c} -U_{j1}
                      \\ Y_{\tau} U_{j2} \end{array} \right) .
\end{eqnarray}
The ${\tilde q}_i$-$q$-${\tilde \chi}^0_k$ couplings are given by
\begin{equation}
a^{\tilde q}_{ik} = {\cal{R}}^{\tilde q}_{in} {\cal{A}}^{f}_{kn}, \hspace{5mm}
b^{\tilde q}_{ik} = {\cal{R}}^{\tilde q}_{in} {\cal{B}}^{f}_{kn}
\end{equation}
with
\begin{equation}
 {\cal{A}}^{f}_{k} = \left( \begin{array}{c} f^f_{Lk}
                      \\ h^f_{Rk} \end{array} \right), \hspace{5mm}
 {\cal{B}}^{f}_{k} = \left( \begin{array}{c} h^f_{Lk}
                      \\ f^f_{Rk} \end{array} \right),
\end{equation}
and
\begin{eqnarray}
\begin{array}{l}
h^t_{Lk} =  Y_t \big( \sin \beta N_{k3} - \cos \beta N_{k4} \big) \\
f^t_{Lk} = \frac{-2 \sqrt{2}}{3} \sin \theta_W N_{k1}
         - \sqrt{2} \big( \frac{1}{2} -
         \frac{2}{3} \sin \theta_Wq \big) \frac{N_{k2}}{\cos \theta_W} \\
h^t_{Rk} =  Y_t \big( \sin \beta N_{k3} - \cos \beta N_{k4} \big) \\
f^t_{Rk} =  \frac{-2\sqrt{2}}{3} \sin \theta_W
              \big( \tan \theta_W N_{k2} - N_{k1} \big)
\end{array}
\end{eqnarray}
\begin{eqnarray}
\begin{array}{l}
h^b_{Lk} = - Y_b \big( \cos \beta N_{k3} + \sin \beta N_{k4} \big) \\
f^b_{Lk} = \frac{\sqrt{2}}{3} \sin \theta_W N_{k1} + \sqrt{2} \big(
    \frac{1}{2} -
         \frac{1}{3} \sin \theta_Wq \big) \frac{N_{k2}}{\cos \theta_W} \\
h^b_{Rk} = - Y_b \big( \cos \beta N_{k3} + \sin \beta N_{k4} \big) \\
f^b_{Rk} =  \frac{\sqrt{2}}{3} \sin \theta_W
            \big( \tan \theta_W N_{k2} - N_{k1} \big)
\end{array}
\end{eqnarray}
where $N_{ij}$ is the mixing matrix of the neutralinos \cite{Bartl89a}.
The couplings $\overline{\tilde t}_i$-${\tilde b}_j$-$W^+$ read
\begin{eqnarray}
A^W_{{\tilde t}_i{\tilde b}_j} = (A^W_{{\tilde b}_i{\tilde t}_j})^T
= \frac1{\sqrt2 \,} \left( \begin{array}{rr}
 \cos \theta_{\tilde b} \cos \theta_{\tilde t} & - \sin \theta_{\tilde b}
            \cos \theta_{\tilde t} \\
 -\cos \theta_{\tilde b} \sin \theta_{\tilde t} & \sin \theta_{\tilde b}
               \sin \theta_{\tilde t} \end{array} \right) \, .
\label{eq:coupstoWsbo}
\end{eqnarray}
The couplings $\overline{\tilde t}_i$-${\tilde b}_j$-$H^+$ are given by
\begin{eqnarray} \hspace*{-6mm}
C^H_{{\tilde t}_i{\tilde b}_j} & = &
            (C^H_{{\tilde b}_i{\tilde t}_j})^T \nonumber \\
 & & \hspace{-12mm} =
\frac1{\sqrt2 \, m_W} {\cal{R}}^{{\tilde t}}
\left( \begin{array}{cc}
   m^2_b \tan \beta + m^2_t \cot \beta - m^2_W \sin(2 \beta) &
                m_b (A_b \tan \beta + \mu) \\
   m_t (A_t \cot \beta + \mu)  & 2 m_b m_t / \sin(2 \beta)
     \end{array}  \right)
\left(  {\cal{R}}^{{\tilde b}} \right)^{\dagger}. \nonumber \\
\label{eq:coupstoHsbo}
\end{eqnarray}
The $W^+$-${\tilde \chi}^-_j$-${\tilde \chi}^0_k$ couplings read:
\begin{eqnarray}
 O^L_{kj}{}' & = & \frac{V_{j2}}{\sqrt2 \,}
    \left( \sin \beta N_{k3} - \cos \beta N_{k4} \right)
  + V_{j1} \left( \sin \theta_W N_{k1} + \cos \theta_W N_{k2} \right) \, , \\
 O^R_{kj}{}' & = & \frac{U_{j2}}{\sqrt2 \,}
    \left( \cos \beta N_{k3} + \sin \beta N_{k4} \right)
  + U_{j1} \left( \sin \theta_W N_{k1} + \cos \theta_W N_{k2} \right) \, .
\end{eqnarray}
The $H^+$-${\tilde \chi}^-_j$-${\tilde \chi}^0_k$ couplings are given by:
\begin{eqnarray}
 Q^L_{kj}{}' & = & \cos \beta \bigg[ V_{j1}
    \left( \cos \beta N_{k4} - \sin \beta N_{k3} \right) \nonumber \\
 & & \hspace{10mm} + \frac{V_{j2}}{\sqrt2 \,}
                   \left( 2 \, \sin \theta_W N_{k1}
      + \left( \cos \theta_W  - \sin \theta_W \tan \theta_W \right)
                N_{k2} \right) \bigg] \, , \\
 Q^R_{kj}{}' & = & \sin \beta \bigg[ U_{j1}
    \left( \cos \beta N_{k3} + \sin \beta N_{k4} \right) \nonumber \\
 & & \hspace{12mm} - \frac{U_{j2}}{\sqrt2 \,}
               \left( 2 \, \sin \theta_W N_{k1}
      + \left( \cos \theta_W  - \sin \theta_W \tan \theta_W \right)
           N_{k2} \right) \bigg] \, .
\end{eqnarray}

\end{appendix}



\begin{table}
\begin{tabular}{llll}
Input: & $\tan \beta = 3$ & $\mu = 530$ GeV & $M = 270$ GeV \\
 & $M_{\tilde D}=370$ GeV & $M_{\tilde Q}=340$ GeV & $A_b=150$ GeV \\
& $M_{\tilde E}=210$ GeV & $M_{\tilde L}=210$ GeV & $A_\tau=150$ GeV \\ 
& $m_{{\tilde t}_1}=250$ GeV & $\cos \theta_{\tilde t}=0.6$ & \\
Calculated & $m_{{\tilde \chi}^0_1}=130$ &
       $m_{{\tilde \chi}^+_1}=253$ & $m_{{\tilde \chi}^+_2}=550$  \\
 &  $m_{{\tilde b}_1}=342$ GeV & $m_{{\tilde b}_2}=372$ GeV & 
      $\cos \theta_{\tilde b}=0.98$ \\
  &  $m_{{\tilde \tau}_1}=209$ GeV &
      $m_{{\tilde \tau}_2}=217$ GeV &  $\cos \theta_{\tilde \tau}=0.68$ \\ 
 & $m_{{\tilde e}_L}=213$ GeV &
      $m_{{\tilde \nu}_e}=m_{{\tilde \nu}_\tau}=204$ GeV & 
\end{tabular}
\caption[]{Input parameters and resulting quantities used in Fig.~2 and 3.}
\label{tabbrst3cosa}
\end{table}

\begin{table}
\begin{tabular}{llll}
Input: & $\tan \beta = 3$ & $\mu = 750$ GeV & $M = 380$ GeV \\
 & $M_{\tilde D}=550$ GeV & $M_{\tilde Q}=500$ GeV & $A_b=400$ GeV \\
& $M_{\tilde E}=275$ GeV & $M_{\tilde L}=275$ GeV & $A_\tau=400$ GeV \\ 
& $m_{{\tilde t}_1}=350$ GeV & $\cos \theta_{\tilde t}=0.7$
   & $m_{A^0}=110$ GeV \\
Calculated & $m_{{\tilde \chi}^0_1}=186$ &
       $m_{{\tilde \chi}^+_1}=368$ & $m_{{\tilde \chi}^+_2}=764$  \\
 &  $m_{{\tilde b}_1}=502$ GeV & $m_{{\tilde b}_2}=551$ GeV & 
      $\cos \theta_{\tilde b}=0.99$ \\
  &  $m_{{\tilde \tau}_1}=274$ GeV &
      $m_{{\tilde \tau}_2}=281$ GeV &  $\cos \theta_{\tilde \tau}=0.69$ \\ 
 & $m_{{\tilde e}_L}=278$ GeV &
      $m_{{\tilde \nu}_e}=m_{{\tilde \nu}_\tau}=270$ GeV & $m_{H^+}=136$ GeV
\end{tabular}
\caption[]{Input parameters and resulting quantities used in Fig.~4 and 5.}
\label{tabbrst3cospmu}
\end{table}


\end{document}